\documentclass[aps,twocolumn,superscriptaddress,amsmath,amssymb]{revtex4-1}
\usepackage{graphicx}
\usepackage{epsfig}
\usepackage{epstopdf}
\usepackage{bm}
\usepackage[usenames,dvipsnames]{color}

\begin{document}

\title{Role of surface termination in the metal-insulator transition of V$_2$O$_3$(0001) ultrathin films}
\author{Asish K. Kundu}
\email{asishkumar2008@gmail.com}
\affiliation{Surface Physics and Material Science Division, Saha Institute of Nuclear Physics, HBNI, 1/AF Bidhannagar, Kolkata 700064, India}
\affiliation{Condensed Matter Physics and Materials Science Department, Brookhaven National Laboratory, Upton, New York 11973, USA}
\author{Sukanta Barman}
\affiliation{Surface Physics and Material Science Division, Saha Institute of Nuclear Physics, HBNI, 1/AF Bidhannagar, Kolkata 700064, India}
\affiliation{Department of Physics, Raja Peary Mohan College, Uttarpara, Hooghly 712258, India}
\author{Krishnakumar S. R. Menon}
\email{krishna.menon@saha.ac.in}
\affiliation{Surface Physics and Material Science Division, Saha Institute of Nuclear Physics, HBNI, 1/AF Bidhannagar, Kolkata 700064, India}

\begin{abstract}

Surface termination is known to play an important role in determining the physical properties of materials. It is crucial to know how surface termination affects the metal-insulator transition (MIT) of V$_2$O$_3$ films for both fundamental understanding and its applications. By changing growth parameters, we achieved a variety of surface terminations in V$_2$O$_3$ films that are characterized by low energy electron diffraction (LEED) and photoemission spectroscopy techniques. Depending upon the terminations, our results show MIT can be partially or fully suppressed near the surface region due to the different filling of the electrons at the surface and sub-surface layers and change of screening length compared to the bulk. Across MIT, a strong redistribution of spectral weight and its transfer from high-to-low binding energy regime is observed in a wide-energy-scale. Our results show total spectral weight in the low-energy regime is not conserved across MIT, indicating a breakdown of `sum rules of spectral weight', a signature of a strongly correlated system. Such change in spectral weight is possibly linked to the change in hybridization, lattice volume ({\it i.e.,} effective carrier density), and spin degree of freedom in the system that happens across MIT. We find that MIT in this system is strongly correlation-driven where the electron-electron interactions play a pivotal role. Moreover, our results provide a better insight in understanding the electronic structure of strongly correlated systems and highlight the importance of accounting surface effects during interpretation of the physical property data mainly using surface sensitive probes, such as surface resistivity.

\end{abstract}
\maketitle
\section{Introduction}
Study of the metal-insulator transition in strongly correlated transition metal oxides such as Ti$_2$O$_3$, VO$_2$, V$_2$O$_3$, and NbO$_2$ are still an active field of research for the understanding and tuning of MIT for their potential applications in intelligent windows and field-effect transistors \cite{zhou2013vo2,yamamoto2019gate,brockman2011increased}. These materials are known to undergo a temperature-dependent MIT in concurrence with a structural transition \cite{mott1974metal,PhysRevMaterials.3.124602}. In addition to the structural transition, the magnetic transition also occur simultaneously in V$_2$O$_3$ and is found to play an important role in MIT \cite{trastoy2018magnetic}. Numerous studies show, a small change of the crystal structure, by Cr or Ti substitution \cite{kuwamoto1980electrical,mcwhan1969mott,shin1995photoemission,lupi2010microscopic} or applying pressure \cite{limelette2003universality} can have a major effect on the MIT of bulk V$_2$O$_3$. Other factors such as surface reconstruction and lattice defects can also affect the MIT of these materials \cite{wickramaratne2019role,panaccione2007bulk}. The breakdown of Mott physics at the surface of VO$_2$  thin films is reported by Wahila {\it et al.} \cite{wahila2020breakdown} due to the surface reconstructions/terminations. For V$_2$O$_3$ thin-films, various types of surface termination have been reported by several groups \cite{dupuis2003v2o3,pfuner2005metal,feiten2015surface1} which can make the understanding of the physics of MIT even more complicated.

At room temperature, V$_2$O$_3$ is a paramagnetic metal (PM) with a corundum structure while at low temperature, below about $\sim$150-160 K, it undergoes a transition to an antiferromagnetic insulating (AFI) phase accompanied by a structural change from the trigonal (corundum) to a monoclinic one with a 1.4\% volume increase. Despite the extensive studies of MIT on V$_2$O$_3$ single crystal and polycrystalline powder, contradictory results persist in thin-films \cite{luo2004thickness,dillemans2014evidence,grygiel2007thickness}. Recently, Luo {\it et al.} \cite{luo2004thickness} have reported a thickness-dependent metal-insulator transition of V$_2$O$_3$ thin-film grown on Al$_2$O$_3$(0001) substrate. They argue that the MIT with decreasing film thickness is due to the increase of the $c/a$ lattice parameter ratio because of the substrate-induced strain in the films. Their results show the thicker films (20 nm) remain in the metallic phase down to 4 K. In contrast, Dillemans  {\it et al.} \cite{dillemans2014evidence} have shown, all their films with thickness ranging from 4$-$73 nm grown on the same substrate (Al$_2$O$_3$) undergo temperature-dependent MIT. Further, x-ray diffraction (XRD) studies in these films (thickness range of 10-100 nm) do not show any systematic change of lattice parameters \cite{grygiel2007thickness}. Schuler {\it et al.} \cite{schuler1997influence} have addressed an impact of the synthesis conditions on MIT and growth modes of the films. All these together suggest the absence of MIT observed in ref. \cite{luo2004thickness,grygiel2007thickness} can not be explained only by considering c/a ratio change.

It is known that V$_2$O$_3$(0001) films generally contains various types of surface termination which changes upon growth conditions \cite{feiten2015surface,feiten2015surface1,dupuis2003v2o3,pfuner2005metal}. Films grown at ultra-high vacuum (UHV) in optimal growth condition favours the formation of vanadyl (V=O) terminated surface \cite{feiten2015surface,feiten2015surface1}. Further, annealing the V=O terminated films at higher oxygen partial pressure gives rise to a $(\sqrt{3}\times\sqrt{3})R30^o$ surface due to the removal of 1/3 or 2/3 of the V=O groups from the surface layer \cite{dupuis2003v2o3,pfuner2005metal}. From the theoretical phase diagram \cite{feiten2015surface} of oxygen chemical potential vs. growth temperature, complete removal of V=O groups is also possible at even higher oxygen partial pressure and temperature which should reflect as a reconstructed O$_3$ terminated (rec-O$_3$) surface layer \cite{feiten2015surface,kresse2004v2o3}. The structural model for all these surface termination as obtained from density function theory (DFT) \cite{feiten2015surface} are shown in figure~\ref{Fig1}(e). Depending upon the terminations, the electronic structure at the surface/sub-surface is expected to deviate from its bulk and its impact should reflect on the electronic properties and MIT. Pfuner {\it et al.} \cite{pfuner2005metal} have reported suppression of MIT for the $(\sqrt{3}\times\sqrt{3})R30^o$ terminated V$_2$O$_3$(0001) surface. It appears that during the interpretation of the experimental data such as surface resistivity, an account of the surface termination effect, in addition to the substrate-induced strain, is essential to get the complete picture of MIT. However, it is not typically discussed how breaking crystal symmetry via surface termination affects the MIT and detailed systematic spectroscopic studies are barely available in the literature. In general, in strongly correlated systems, the change of hybridization, lattice volume (electron density), spin degree of freedom can show unusual spectral evolution mainly in the low-energy scale \cite{meinders1993spectral,kohno2019emergence,rozenberg1996transfer}, however, there is still little understanding how MIT in V$_2$O$_3$ alters the low-energy physics.

 Here, we study the effect of surface termination on the electronic structure of V$_2$O$_3$ films and its impact on MIT. To achieve this goal, well ordered V$_2$O$_3$(0001) surface has been grown on Ag(111) substrate with different oxygen partial pressures and substrate temperatures, following post-growth procedures. Our results show, at optimal growth condition, the V$_2$O$_3$(0001) surface is terminated by the V=O groups while annealing the film at higher oxygen partial pressure and temperature results in the $(\sqrt{3}\times\sqrt{3})R30^o$ surface structure due to the removal of V=O groups. Films directly grown at higher oxygen partial pressures and temperatures results in a nearly rec-O$_3$-like surface. Partial to full suppression of MIT is observed by going from V=O to $(\sqrt{3}\times\sqrt{3})R30^o$ and rec-O$_3$-like surfaces. We also show that MIT in this system is strongly correlation driven and it is associated with a strong redistribution of orbital occupancy in the much wider energy scale than previously thought. Further, we discuss the possible origin of unusual spectral weight enhancement in the low-energy-scale that occurs when the system is cooled below MIT.

\section{Results and discussion}
\subsection{Surface structure of V$_2$O$_3$ films}
 \begin{figure*}[ht!]
\centering
\includegraphics[width=12cm]{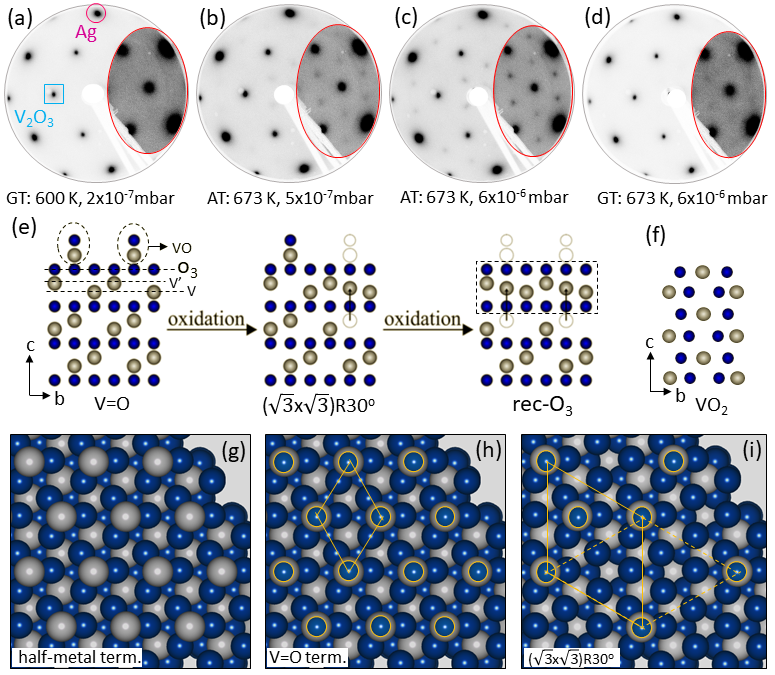}
\caption {LEED pattern of V$_2$O$_3$ films grown on Ag(111) at various substrate temperatures and oxygen partial pressures, along with structural models. (a) Films grew at 600 K in presence of $P(O_2)$= 2$\times$10$^{-7}$ mbar followed by UHV annealing at 773 K. The position of first-order diffraction spots from Ag(111) (circle) and V$_2$O$_3$ (square) are marked. Images with enhanced contrast and brightness are shown inside oval (red). (b) and (c) Post annealing of the film at 673 K, $P(O_2)$= 5$\times$10$^{-7}$ mbar and 673 K, $P(O_2)$= 6$\times$10$^{-6}$ mbar, respectively. (d) Films directly grown at 673 K, $P(O_2)$= 6$\times$10$^{-6}$ mbar. Growth temperatures and post-annealing temperatures are denoted as GT and AT, respectively. (e) Side-view of the structural model (gray (V) and blue (O)) for the vanadyl (V=O), $(\sqrt{3}\times\sqrt{3})R30^o$, and rec-O$_3$ terminated V$_2$O$_3$(0001) surface as obtained from DFT \cite{feiten2015surface}. V=O groups are enclosed by the dotted ovals. Three types of bulk termination can be realized by terminating the surface at V (half-metal), V' (full-metal), and O$_3$ (bulk-O$_3$) layers. Dotted rectangle encloses trilayer structure on top of half-metal terminated V$_2$O$_3$ surface. Trylayer has VO$_2$ stoichiometry but structurally different than the rutile VO$_2$(100) (f). (g)-(i) Top-view of the half-metal, V=O and $(\sqrt{3}\times\sqrt{3})R30^o$ terminated surface, respectively. Vanadyl oxygens are enclosed by the yellow circles. Rhombus in (h) represents the surface unit-cell of V=O terminated surface.  Rhombus with solid and dotted lines in (i) represents the surface unit cell corresponding to the removal of 1/3 and 2/3 of V=O groups, respectively.}\label{Fig1}
\end{figure*}

Figure~\ref{Fig1}(a) shows the LEED image of 30 MLE V$_2$O$_3$ films on Ag(111), grown at a substrate temperature of 600 K in presence of $P(O_2)$= 2$\times$10$^{-7}$ mbar followed by UHV annealing at 773 K. The position of the first order diffraction spots from Ag(111) (before deposition of films) and V$_2$O$_3$ are marked. LEED pattern shows $(\sqrt{3}\times\sqrt{3})R30^o$ structure {\it w.r.t} Ag(111) substrate. It is expected as the lattice parameter of the V$_2$O$_3$(0001) surface (4.95 {\AA}) is very close to the $\sqrt{3}$ times of Ag(111) lattice parameter ($\sqrt{3}\times$2.89=5.0 {\AA}) with only 1\% lattice mismatch, favouring the growth of epitaxial V$_2$O$_3$(0001) films. This structure can be thought as (1$\times$1) structure {\it w.r.t} the V$_2$O$_3$(0001) basis vectors. Throughout the paper, we will call this structure as (1$\times$1) structure for the simplicity of our discussions. The (1$\times$1) structure have been also observed by previous studies of V$_2$O$_3$ films on Au(111) and W(110) substrate at similar growth conditions which was regarded as a V=O terminated V$_2$O$_3$(0001) surface \cite{dupuis2003v2o3,pfuner2005metal}. The V=O terminated surface may be thought as an ideal half-metal terminated surface (figure~\ref{Fig1}(g)) of V$_2$O$_3$(0001) with one additional oxygen atom bonded on top of each outermost V atom (figure~\ref{Fig1}(h)). Annealing the V=O terminated surface in presence of oxygen shows additional spots (superstructures) that form $(\sqrt{3}\times\sqrt{3})R30^o$ structure {\it w.r.t.} (1$\times$1)-V$_2$O$_3$ as shown in figure~\ref{Fig1}(b). These superstructure spots gain intensities upon further annealing the sample at higher oxygen pressure as can be seen in
figure~\ref{Fig1}(c). The $(\sqrt{3}\times\sqrt{3})R30^o$ structure could originate by removing the 1/3 or 2/3 amount of V=O groups from the V=O terminated surface \cite{schoiswohl2004v2o3,feiten2015surface1}. The surface atomic configuration is shown in figure~\ref{Fig1}(i), where the rhombus with solid and dotted lines represent the surface unit cell corresponding to the removal of 1/3 and 2/3 of V=O groups, respectively. Relatively lower intense and broad superstructure spots in figure~\ref{Fig1}(b) compare to figure~\ref{Fig1}(c) may indicate that the removal of V=O groups is less uniform and lower in numbers in figure~\ref{Fig1}(b) than in figure~\ref{Fig1}(c). Similar behavior has been reported by Schoiswohl {\it et al.} \cite{schoiswohl2004v2o3} using combined scanning tunneling microscopy (STM) and LEED study.

Figure~\ref{Fig1}(d) shows the LEED pattern of films that has been directly grown at higher temperature and oxygen partial pressure (673 K, $P(O_2)$= 6$\times$10$^{-6}$ mbar). The LEED pattern shows mainly (1$\times$1) structure but additional broad spots with extremely faint intensities can be visible at higher brightness/contrast (inside oval) that form $(\sqrt{3}\times\sqrt{3})R30^o$ structure. Feiten {\it et al.} \cite{feiten2015surface}, have pointed out that both V=O and rec-O$_3$ terminated surface exhibit identical reflex patterns and indicated that the LEED-IV measurements might be helpful to distinguish them. Further, DFT calculation \cite{feiten2015surface} also predicts that the rec-O$_3$ surface termination favors at higher temperature and oxygen partial pressure (see the structural model in figure~\ref{Fig1}(e)). Thus, the presence of extremely low intense $(\sqrt{3}\times\sqrt{3})R30^o$ structure could mean that only locally 1/3 or 2/3 amount of  V=O groups are missing or left. In other words, this surface is either very close to the V=O terminated or rec-O$_3$ terminated surface.

 \begin{figure}[ht!]
\centering
\includegraphics[width=8.5cm]{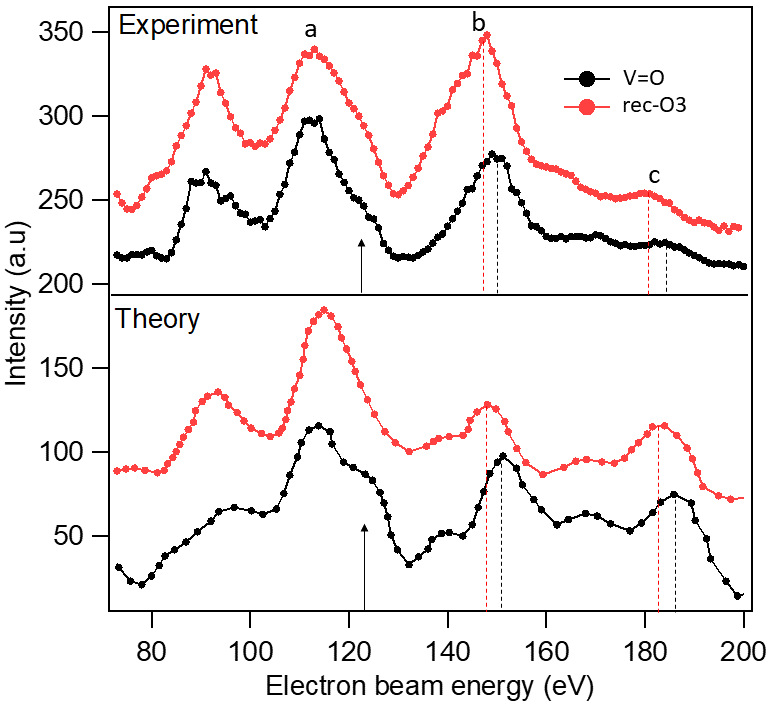}
\caption{Comparison of experimental LEED-IV curves with theory \cite{feiten2015surface} for V=O (GT: 600 K, $P(O_2)$= 2$\times$10$^{-7}$ mbar followed by UHV annealing at 773 K) and rec-O$_3$ (GT: 673 K, $P(O_2)$= 6$\times$10$^{-6}$ mbar) terminated surface. Experiment (top panel) and theory (bottom panel). The peak marked by `a' shows a prominent two-peak structure for V=O terminated surface, consistent with theory. Relative to the rec-O$_3$ termination, the `b' and `c' features shifted towards higher energy for the V=O terminated surface. Arrows at 124 eV indicate the appearance of a peak for the V=O terminated surface which is absent for rec-O$_3$. }\label{Fig2}
\end{figure}

To get more insight into the surface structure, we have performed experimental LEED-IV measurements and compared them with the reported theoretical results \cite{feiten2015surface}. The top panel in figure~\ref{Fig2} shows the experimental LEED-IV curves extracted from the first-order diffraction spots of V$_2$O$_3$ corresponding to the figure~\ref{Fig1}(a) (black) and figure~\ref{Fig1}(d) (red), respectively. The bottom panel in figure~\ref{Fig2} shows the simulated LEED-IV curves for the V=O and rec-O$_3$ terminated surfaces \cite{feiten2015surface}. Between the two experimental curves, there are some significant differences can be observed such as black spectra shows prominent two peak structure for the peak `a', while peaks `b' and `c' shifted towards the higher energy compared to the red spectra. By comparing with the theoretical spectra, it appears that the position of peaks in black experimental spectra better match with the V=O terminated while the red spectra with the rec-O$_3$ terminated surface. Although it is not perfectly rec-O$_3$ terminated surface as in that case it would show perfectly ($1\times1$) LEED pattern \cite{feiten2015surface}, but we will call it rec-O$_3$-like due to its similarities with the LEED-IV curves with the rec-O$_3$ terminated surface.  We note that the position of peaks in the experimental I-V curves are in agreement with the theoretical curves whereas peak intensities differ. The experimental curves are not normalized by the incident electron beam-current that could produce the observed intensity differences. Unfortunately, we could not correct it as there were some technical issues with the beam-current reader during I-V data acquisition. The normalization can only change the intensity of the peaks, not the peak features or their position. Thus, it will not affect the main claims of our study.

\subsection{Electronic structure and MIT of V$_2$O$_3$ films}

 \begin{figure}[ht!]
\centering
\includegraphics[width=8.5cm]{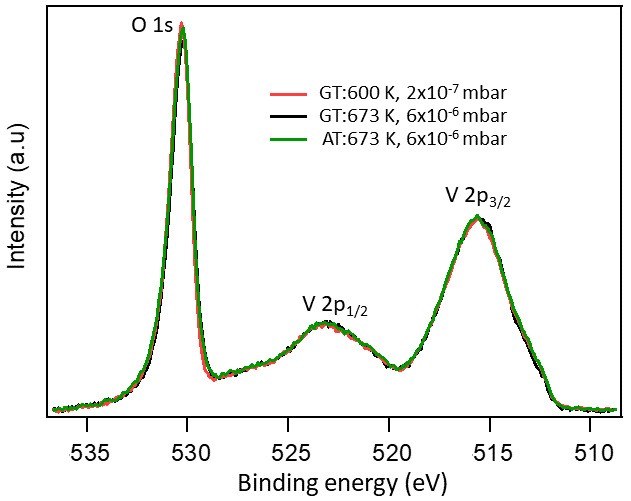}
\caption{XPS core-levels of O $1s$, V 2$p_{1/2}$ and V 2$p_{3/2}$ for V=O (GT: 600 K, 2$\times$10$^{-7}$ mbar), $(\sqrt{3}\times\sqrt{3})R30^o$ (AT: 673 K, 6$\times$10$^{-6}$ mbar) and rec-O$_3$ (GT: 673 K, 6$\times$10$^{-6}$ mbar) terminated surfaces. All the spectra are Shirley background subtracted.}\label{Fig3}
\end{figure}
Understanding the chemical composition of these films is essential as vanadium may exist at different oxidation states (V$^{+2}$ to V$^{+5}$ ) depending upon the growth conditions. The most common way to directly access the chemical state of materials is to perform the XPS measurements. Figure~\ref{Fig3} shows the O $1s$, V 2{\it p}$_{1/2}$ and V 2{\it p}$_{3/2}$ core-levels of V=O, $(\sqrt{3}\times\sqrt{3})R30^o$, and rec-O$_3$ terminated films. Despite the different surface structures observed in LEED, all the XPS spectra show very similar spectral features, indicating that they have a similar chemical composition in the bulk. We note that depending upon the used photon energy, XPS can be surface or bulk sensitive as its probing depth varies depending upon materials, the kinetic energy, and photoemission angle of the photoelectrons being measured \cite{brundle2020x}. Here, the kinetic energy of the V 2{\it p} photoelectrons measured using Al K$_\alpha$ (hv=1486.6 eV) is $\sim$ 965 eV that corresponds to $\lambda$ $\sim$ 2.06 nm and probing depth 3$\lambda$ $\sim$ 6.2 nm. Where $\lambda$ is the attenuation length (mean free path) of photoelectrons. As the probing depth, $\sim$ 6.2 nm is comparable to the film thickness $\sim$ 6.5$-$7 nm, thus the surface effect is expected to be suppressed here.

 \begin{figure*}[ht!]
\centering
\includegraphics[width=16cm]{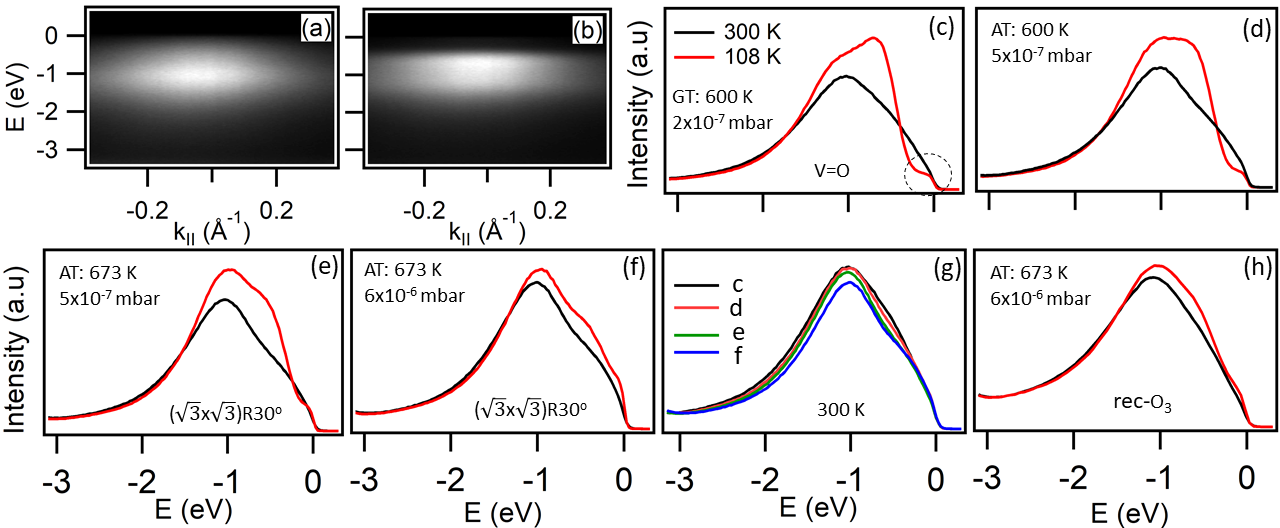}
\caption{ARPES intensity plot and EDCs at $\bar{\Gamma}$ for various V$_2$O$_3$ films (30 MLE) below (red) and above (black) MIT. (a) and (b) are the ARPES intensity plot for the V=O terminated surface at 300 K and 108 K, respectively. (c) EDCs (at $\bar{\Gamma}$) as obtained from (a) and (b) are plotted together. Inside dotted circle a Fermi edge-like feature (red spectra) can be seen. (d) Annealing the V=O terminated film at 600 K, $P(O_2)$= 5$\times$10$^{-7}$ mbar for 10 min, (e) Annealing of the film (d) at 673 K, $P(O_2)$= 5$\times$10$^{-7}$ mbar for 10 min. (f) Annealing of the film (e) at 673 K, $P(O_2)$= 6$\times$10$^{-6}$ mbar for 10 min. (g) EDCs at 300 K for samples (c)-(f) plotted together. (h) Films directly grown at 673 K, $P(O_2)$= 6$\times$10$^{-6}$ mbar (rec-O$_3$). All the data were taken using He I$_\alpha$ (21.2 eV) photons. }\label{Fig4}
\end{figure*}
XPS spectra were fitted (not shown) to extract the exact peak position and their full width at half maximum (FWHM). The details of fitting can be found elsewhere in our previous study \cite{kundu2017thickness}. The obtained peak position of O $1s$, V 2{\it p}$_{1/2}$ and V 2{\it p}$_{3/2}$ core-levels are 530.2 eV, 522.85 eV and 515.6 eV, respectively. In the case of vanadium oxide, the most common way to identify the exact phase is to measure the energy separation ($\Delta$) between the O $1s$ and V 2p$_{3/2}$ core-levels and their FWHM \cite{mendialdua1995xps}. Here, the obtained value of $\Delta$ is 14.7 eV and FWHM of O $1s$ and V 2p$_{3/2}$ peaks are 1.3 eV and 4.05 eV, respectively. All these values lie in the stoichiometric region of V$_2$O$_3$ \cite{kundu2017thickness,sawatzky1979x}. The overall spectral line-shape is also in good agreement with the V$_2$O$_3$ films \cite{kundu2017thickness,sawatzky1979x,kamakura2004hard}. These results are in contrast with the DFT predictions \cite{feiten2015surface}, as the bulk phase of rec-O$_3$ terminated films is expected to be already in the region of V$_2$O$_5$ stoichiometry. This inconsistency may come from the overestimation of the Coulomb correlation effects in DFT calculation and the choice of exchange-correlation functional. In our experimental growth conditions, it may be possible that the formation of a higher oxidation state in bulk is just prevented by kinetic limitations.

Further, it is important to know how surface termination affects the electronic structure and the MIT. Figure~\ref{Fig4}(a) and (b) show the ARPES intensity plots within the V $3d$ region of V=O terminated V$_2$O$_3$ films, above and below MIT temperatures, respectively. The energy distribution curves (EDCs) obtained from figure~\ref{Fig4}(a) and (b) are shown in figure~\ref{Fig4}(c). In pure V$_2$O$_3$, V$^{3+}$ has two $3d$ electrons, occupying the triply degenerate $t_{2g}$ (a$_{1g}$, e$_g^\pi$) orbitals, split into a lower e$_g^\pi$ doublet and an upper $a_{1g}$ singlet in the trigonal crystal field (CF) of PM phase. In the low-temperature monoclinic phase, the additional low-symmetry CF component lifts e$_g^\pi$ degeneracy. Moreover, EDCs of V $3d$ region is dominated by  e$_g^\pi$ and a$_{1g}$ orbital. In the paramagnetic phase (figure~\ref{Fig4}(a)), the bands at $\sim$ -1.1 eV and $\sim$ -0.4 eV are arises from the lower Hubbard band (LHB) and quasiparticle peak (QP), respectively, according to the previous studies \cite{mo2003prominent,rodolakis2009quasiparticles}. The QP is broad and less intense due to the lower photon energy (21.2 eV) used in our experiment, where we are extremely sensitive to the surface. Previous studies also show less intense QP at the surface than bulk due to the enhanced correlation at the surface region relative to the bulk \cite{mo2003prominent}. This is also in line with the DFT+ dynamical mean-field theory (DMFT) results,  where it has been shown that a dead layer forms below the surface of V$_2$O$_3$, where quasiparticles are exponentially suppressed \cite{borghi2009surface}.

In figure~\ref{Fig4}(c), a significant change in spectral weight near Fermi level (E$_F$) and its shifts towards the higher binding energy is observed when the sample is cooled below the MIT temperature. The observed behavior is mainly caused by the change in orbital occupancy in a$_{1g}$ and e$_g^\pi$ orbital of V 3{\it d} across MIT, as observed in x-ray absorption spectroscopy (XAS) measurements \cite{park2000spin}. By lowering the temperature from 300 K (PM) to 108 K (AFI),  it seems like the spectral weight is transferred from QP to the LHB side and the QP and LHB are shifted by $\sim$ 200 meV and $\sim$ 80 meV, respectively, towards higher binding energy. Total width (FWHM) of the V $3d$ region (and LHB) also changes from 1.85 eV (1.4 eV) to 1.54 eV (1.24 eV) by going from PM to AFI phase. These values are obtained from fitting the spectra (see supplementary material, figure S1). According to the Hubbard model, decrease of LHB width ($W$) and its shift towards the higher binding energy, implies that the on-site Coulomb interactions ($U$), as well as correlation ($U/W$) of the system, gets enhanced in the AFI phase compared to the PM phase. This suggests, MIT in V$_2$O$_3$ is strongly correlation driven, where the electron-electron correlation plays an important role. Although the Hubbard model ignores the degeneracy of bands, which is crucial for understanding the magnetic structures \cite{castellani1978magnetic}, it seems well captures the interplay between the electron-electron interactions and kinetic energy. The change of $U$ and $W$ across the phase transition are most likely due to a change of lattice parameter (change the hopping parameters) and effective screening length. Additionally, in the AFI phase, the scattering rate is also found to decrease which might sharpen the bandwidth as well \cite{deng2014shining}. From PM to AFI transition, an increase of $U$ and decrease of $W$ values are also reported by Rozenberg {\it et al.} \cite{rozenberg1995optical} using optical measurements, similar to our observation.

However, it is not clear whether the QP band is totally absent in the AFM phase or it just shifts towards the higher energy with a redistribution of spectral weight. Across MIT, similar spectral features like our results were also reported for the V=O terminated V$_2$O$_3$ films \cite{pfuner2005metal} and bulk single crystals \cite{mo2006photoemission}. This apparent shift of the PM phase spectrum across the MIT has been also found in a recent full-orbital DMFT calculation \cite{anisimov2005full}, although the actual shape and peak position of the theoretical spectrum has different from the experimental spectra. Other DFT+DMFT calculations \cite{lechermann2018uncovering,leonov2015metal} incorporating full charge self-consistency were unable to capture the complex `two-peak structure of the LHB in the insulating (AFI) phase. We note that all these calculations were performed without accounting the exact spin-structure of the AFI phase of V$_2$O$_3$ and some of them only account the PM crystal structure, which may produce the observed discrepancy between theory and experiment. Trastoy {\it et al.} have also shown the importance of magnetic structure in MIT of V$_2$O$_3$ using resistivity and magnetoresistance measurements, combined with the theoretical calculations \cite{trastoy2018magnetic}.

Besides, in figure~\ref{Fig4}(c), we observe a small Fermi-edge-like feature (enclosed by a dotted circle) at E$_F$ in the 108 K spectra which hinders the opening of a full energy gap at E$_F$. This feature is not resolved in the 300 K spectra as it masks by the QP peak. By comparing with the theoretical layer-resolved density of state calculation \cite{kresse2004v2o3}, it appears that this edge-like feature is most probably originating from the V=O groups at the surface of the film. The theoretical data clearly shows the presence of a small density of states (DOS) at E$_F$, which arises from a VO-like surface layer. This feature gains its intensity when measurements were performed using He II$_\alpha$ than He I$_\alpha$ (see supplementary information, figure S2) that strongly suggest its surface-related origin, as He II$_\alpha$ is more surface sensitive than He I$_\alpha$. Furthermore, as VO does not show temperature-dependent MIT thus, the DOS originating from VO should remain at the same energy position both below and above MIT temperatures, which agrees with our results. This peak (V=O related) forms a Fermi-edge-like feature as it is cut by the Fermi Dirac distribution function at E$_F$. We also vary the substrate temperature, a different substrate such as W(110), and oxygen partial pressure in a wide range but this feature was always present with a slight variation of intensity, depending upon the growth condition. According to Schoiswohl {\it et al.} \cite{schoiswohl2004v2o3}, V$_2$O$_3$ films grown at optimal condition followed by UHV annealing produce large-area films with fully covered V=O groups at the surface. In agreement, upon post-annealing the film in UHV, we also observe the intensity of the Fermi-edge-like feature and other V=O-related peaks are enhanced (see supplementary information figure S3). Possibilities of Ag segregation from substrate or formation of microcracks in the film can be ruled out, as no intensity enhancement of Ag-related peaks is observed (figure S3). Our previous study shows strong enhancement of Ag-related peak intensity when Ag segregates on V islands \cite{kundu2018structural}. However, our study can not completely rule out the possibility of the presence of some precursor metallic phase even at 108 K, as the coexistence of microscopic metallic and insulating islands was found in a wide temperature range for Cr-doped V$_2$O$_3$ sample \cite{lupi2010microscopic}. Moreover,  It is very unlikely that the V=O surface layer with its different geometry and electronic structure can support the MIT. Thus, the observation of spectral change across MIT in figure~\ref{Fig4}(c) and opening of a partial energy gap below MIT temperature, clearly suggest that the MIT in the bulk of the V$_2$O$_3$ film is communicated through the V=O surface layer.

Post annealing the V=O terminated films with increasing oxygen partial pressure and temperature are shown in figure~\ref{Fig4}(d)-(f). It can be seen that MIT is progressively weakened and completely suppressed by going from figure~\ref{Fig4}(d) to figure~\ref{Fig4}(f). The complete suppression of MIT is also observed for the rec-O$_3$ like surface as shown in figure~\ref{Fig4}(h). In figure~\ref{Fig4}(e), change of spectral weight across MIT is still observed but without a gap at E$_F$, whereas in figure~\ref{Fig4}(f), the spectral change is minimal and MIT is completely suppressed.  In figure~\ref{Fig4}(g), all the EDCs from the figure~\ref{Fig4}(c) to (f) films taken at 300 K are plotted together. It is evident that with increasing oxygen partial pressure and temperatures, the total area of the V 3$d$ region (0$-$3 eV) decreases about $\sim$ 16\% (from figure~\ref{Fig4}(c) to (f)) which suggests a decrease of 3$d$ electrons from 2 for V$_2$O$_3$ to the lower numbers, near the surface layer. The decrease in electron counts suggests that the surface is getting more oxygen-rich. A similar change of spectral features and decrease of V 3$d$ area are reported for bulk V$_{2-y}$O$_3$ sample \cite{kim1998high}. These suggest that the bulk MIT of V$_2$O$_3$ films are more effectively screen for the $(\sqrt{3}\times\sqrt{3})R30^o$ and rec-O$_3$ like oxygen-rich surfaces than (1$\times$1) V=O terminated surface. Suppression of MIT near the surface region can be understood by considering the structural change which occurs during the formation of rec-O$_3$ and $(\sqrt{3}\times\sqrt{3})R30^o$ surfaces, where the removal of each V=O group exposes three oxygen atoms of the underlying O$_3$ plane (see the structural model in figure~\ref{Fig1}(g) and (i)). To maintain thermodynamical equilibrium, the surface region, including subsurface layers relaxes and undergoes rearrangements (figure~\ref{Fig1}(e)) \cite{schoiswohl2004v2o3}, which strongly deviates from corundum bulk structure, whereas for V=O surface, only the top layer gets affected while subsurface layers preserve their bulk structure. Suppression of MIT is also observed in resistivity measurements for the oxygen-rich sample \cite{trastoy2018enhanced}. Surface termination act as a charge doping (here, hole doping) to the surface layers, thus drastically alter the MIT characteristics by pushing the $d$-filling further or closer to the Mott criterion \cite{lee2019cooperative}.

We note that for rec-O$_3$ and $(\sqrt{3}\times\sqrt{3})R30^o$ (underlying surface below the removed V=O groups) structures, the surface has a stacking sequence -V$_2$O$_3$-V-O$_3$V$_3$O$_3$ which may be thought of as an O$_3$V$_3$O$_3$ trilayer structure with a formal VO$_2$ stoichiometry on top of the single V terminated bulk structure (see figure~\ref{Fig1}(e)).  As VO$_2$ is known to show MIT below 320 K \cite{choi2021correlation}, both the 300 K and 108 K EDCs corresponding to these structures would show insulating behavior or at least the 108 K EDCs, as the underlying structure is still V$_2$O$_3$ which only shows MIT at low temperature. While our data show metallic behavior in both temperatures (figure~\ref{Fig4}(f) and (h)). This behavior suggests that the chemical composition alone cannot determine the MIT properties but a combination of chemical composition and crystal structure does. The structural difference between the trilayer stacking with VO$_2$ composition and the bulk VO$_2$ are shown in figure~\ref{Fig1}(e) and (f), respectively.

 \begin{figure}[ht!]
\centering
\includegraphics[width=8.5cm]{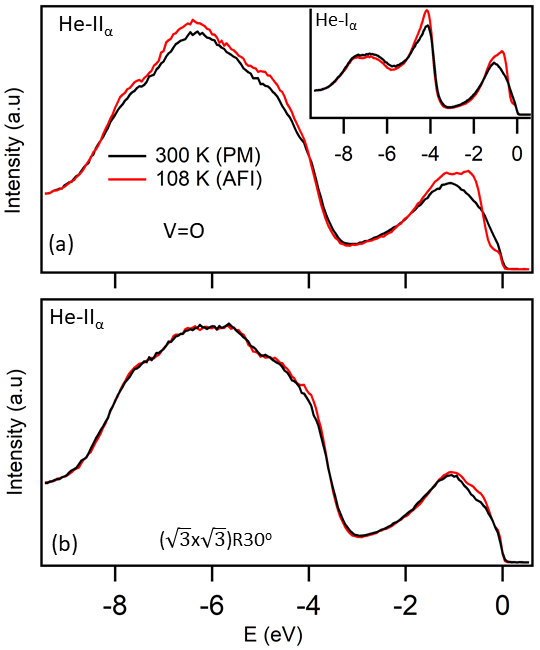}
\caption{ EDCs in a wide energy range around MIT. EDCs using He II$_\alpha$ (40.8 eV) photons for (a) V=O and (b) $(\sqrt{3}\times\sqrt{3})R30^o$ (AT: 673 K, $P(O_2)$= 6$\times$10$^{-6}$ mbar) structure. Inset of (a) shows EDCs of V=O terminated surface using He I$_\alpha$ photons.}\label{Fig5}
\end{figure}

We also note that in figure~\ref{Fig4}(c), the integrated area within the energy ranges, 0 to -3.0 eV is enhanced by 16.5\% in the
AFI (108 K) phase compared to PM (300 K), indicating breakdown of `sum rules of spectral weight' in that energy regime which in turn suggests the presence of strong correlation in the system \cite{meinders1993spectral,kohno2019emergence}.
Theoretically, the breakdown of `sum rules of spectral weight' in the low-energy scale has been found in strongly correlated systems when a change of hybridization and/or electron density are introduced in the system \cite{meinders1993spectral,kohno2019emergence}. A question may arise, what is the source of this extra spectral weight? To get further insight, we have plotted EDCs in a wide energy window as shown in figure~\ref{Fig5}(a). Not only in the lower energy regime, but a strong redistribution of spectral weight is also observed within the whole energy range, 0 to -9.5 eV. Spectra taken at He I$_\alpha$ photons (inset of figure~\ref{Fig5}(a)) also shows similar behavior. The main difference between He I$_\alpha$ and He II$_\alpha$ spectra is that spectral weight within the energy ranges, -5 to -8 eV decreases in the AFI phase for He I$_\alpha$ whereas increases for He II$_\alpha$. The different behavior could be associated with the k$_z$ dependence of spectral weight transfer \cite{kohno2019emergence}. The estimated increase of spectral weight in the AFI phase within the whole measured energy ranges are $\sim$ 3.5\% and $\sim$ 6\% for He I$_\alpha$ and He II$_\alpha$, respectively. Previous photoemission studies performed on the Cr-doped V$_2$O$_3$ sample, between the paramagnetic insulator (PI) and the crossover regime also found spectral weight redistribution within the whole valence band region \cite{mo2004filling}. In contrast, their results show that the V 3{\it d} and O 2{\it p} spectral weight conserved separately within these two regions (PI and crossover). Similar to our results, optics experiment also shows total spectral weight across MIT is not conserved within energy range, 0 to 6 eV above the Fermi energy \cite{qazilbash2008electrodynamics,PhysRevB.85.205113,rozenberg1995optical}. Interestingly, it can be seen that for the oxygen-rich $(\sqrt{3}\times\sqrt{3})R30^o$ surface, the spectral weight change is almost negligible as shown in figure~\ref{Fig5}(b). These results confirming that the spectral weight enhancement and its redistribution are strongly related to the MIT transition.

The change in lattice structure and volume that occurs across MIT may have important consequences in the observed spectral weight renormalization (redistribution), as it alters the hybridization strength between different orbital as well as carrier density (carriers/volume) that strongly affect the orbital occupancy. In general, electronic states (bands) are usually considered invariant regardless of the electron density in a band picture. However, in interacting systems, the spectral-weight distribution changes depending on the electron density, and electronic states can even emerge or disappear \cite{kohno2019emergence,meinders1993spectral}. Furthermore, Kohno {\it et al.,} \cite{kohno2019emergence}, have explicitly shown that the electronic excitations much away from the Fermi level can also become dominant and exhibit significant characteristics due to the change in electron density. Keeping in mind that, as we are not externally doping carriers in the system then the total carriers should be conserved but carrier density may change across MIT due to the change in lattice volume. Thus, one feasible interpretation for an apparent increase of spectral weight within 0 to -9.5 eV regime would be that the transfer of spectral weight occurs from much deeper to the lower binding energy side, due to the redistribution of orbital occupancy in the larger energy regime. It can be clearly seen from the He I$_\alpha$ data (inset of figure~\ref{Fig5}) that the spectral weight decreased in the higher binding energy side whereas increased going towards the Fermi energy. Further, theoretical study considering the antiferromagnetic ordering in V$_2$O$_3$ also find evidence of spectral weight transfer from higher to lower binding energy side compared to the PM phase \cite{rozenberg1996transfer}. All these together suggest that the strong change in spectral weight across MIT is due to the complex interplay between the charge, orbital and spin degree of freedom in this system.

 We also note that MIT of V$_2$O$_3$(0001) films have been studied by several groups by resistivity measurement \cite{luo2004thickness,dillemans2014evidence,grygiel2007thickness,brockman2011increased}. Unlike bulk V$_2$O$_3$ single crystals, in most of the thin-films, resistivity {\it vs.} temperature curve do not show sharp transition  \cite{dillemans2014evidence,grygiel2007thickness,brockman2011increased} and in some cases, the system remains in the metallic phase \cite{luo2004thickness,grygiel2007thickness}. Substrate-induced strain effect in the films has been considered as an origin of the absence of the temperature-driven MIT \cite{luo2004thickness,grygiel2007thickness}. In contrast, Dillemans  {\it et al.} \cite{dillemans2014evidence} have shown that V$_2$O$_3$(0001) films grown on the same substrate (Al$_2$O$_3$) undergo temperature-dependent MIT, questions the validity of strain-induced loss of MIT in this system. Our results highlight, in addition to the epitaxial strain, there should be surface termination effects, and combining these two can better explain the observed resistivity change in a wide temperature scale or suppression/absence of MIT in most of the thin-film systems.

\subsection{Conclusion}
In summary, we have grown ultrathin films of V$_2$O$_3$(0001) with various surface terminations and characterized them using LEED and photoemission spectroscopy techniques. We have shown that depending on the types of termination, MIT at the surface region can be partially or fully suppressed. This suggests high sensitivity of the Mott metal-insulator transition to the local environment, consistent with recent theoretical predictions \cite{lechermann2018uncovering}. We have also shown, upon MIT, the spectral weight is not only shifted from QP to the LHB, but a huge redistribution of spectral weight is observed within the whole valence band. Total spectral weight was found to be not conserved within the energy ranges, 0 to -9.5 eV which indicates that the spectral weight is redistributed in much larger energy, suggesting the system is heavily correlated.  Such enhancement of spectral weight in the low-energy regime is mainly due to the spectral weight transfer from higher to lower binding energy regime due to the change in complex interplay between the charge, orbital and spin degree of freedom in this system that occurs during MIT. We have also shown that in addition to the other effects, taking into account surface termination can better explain the anomaly observed in the resistivity data of V$_2$O$_3$. We expect that similar behavior would be at play for other complex oxides and the interpretation of physical properties data (mainly using surface sensitive probe) requires more attention. Furthermore, the observed complex structure of LHB in our experiment (in the AFI phase) is not reproduced by the currently available theories, including DFT+DMFT. Thus, we believe that our results might help in the further development of the theoretical understanding of MIT in strongly correlated systems.

\section{Experimental}

The Ag(111) substrate was cleaned by repeated cycles of Ar$^{+}$ ion sputtering (600 eV, 1 $\mu$A) for 15 min followed by annealing at 823 K for 20 min until a sharp $p(1\times1)$ LEED pattern was observed. The substrate preparation, growth, and LEED measurements were performed in the preparation chamber with a base pressure of 1$\times$10$^{-10}$ mbar. Vanadium was evaporated from a well-degassed water-cooled e-beam evaporator at a constant rate of 0.3 {\AA}/min in oxygen partial pressure range of {\it P}(O$_2$)= $1\times10$$^{-7}-6\times10$$^{-6}$ mbar for the deposition of the V$_2$O$_3$ film, while during deposition, the substrate temperature was varied between at 473$-$673 K. The rate of V-deposition was calibrated with a water-cooled quartz crystal thickness monitor. One monolayer equivalent (MLE) of vanadium oxide film is defined as the atomic density corresponding to 1 ML vanadium on Ag(001), {\it i.e}, 1.75$\times$10$^{19}$ atoms/m$^2$, under oxygen atmosphere. We have grown 30 MLE films that appear $\sim$ 6.5$-$7 nm as estimated from the attenuation of Ag 3{\it d} core-level intensity. To achieve a high-quality surface, the film was annealed at 673$-$773 K for 10 min in UHV. The sample temperature was measured by a K-type thermocouple, in contact with the Ag crystal. The crystalline quality of the film and the symmetry directions were determined by a four-grid LEED apparatus while a highly-sensitive Peltier-cooled 12-bit CCD camera was used to collect the LEED images. Photoemission spectroscopic measurements, XPS, and ARPES were performed in an analysis chamber having a base pressure better than 8$\times$10$^{-11}$ mbar. ARPES experiments were performed using a combination of VG SCIENTA-R4000WAL electron energy analyzer with a 2D-CCD detector and a high flux GAMMADATA VUV He lamp attached to a VUV monochromator, which has been described in detail elsewhere \cite{mahatha2010angle}. The ARPES measurements were performed using monochromatized ultraviolet He I$_\alpha$(21.2 eV) and He II$_\alpha$ (40.8 eV) resonance lines, while the XPS measurements were performed with Al K$_{\alpha}$ monochromatic X-ray source (1486.6 eV) from VG SCIENTA. At room temperature (RT), the total experimental energy resolution was about 100 meV for the UPS (using He I$_\alpha$ and He II$_\alpha$ photons) while for the monochromatic XPS the energy resolution was set to about 600 meV.
\section{acknowledgement}

The Micro-Nano initiative program of the Department of Atomic Energy (DAE), Government of India, is acknowledged for generous funding and support for this work. During writing the paper, A.K.K receives funding from US Department of Energy, Office of Basic Energy Sciences, contract no. DE-SC0012704. The authors thank T. Valla, A. Samanta, and S. K. Mahatha for the helpful discussions.



\bibliography{Ref}

\begin{thebibliography}{52}%
\makeatletter
\providecommand \@ifxundefined [1]{%
 \@ifx{#1\undefined}
}%
\providecommand \@ifnum [1]{%
 \ifnum #1\expandafter \@firstoftwo
 \else \expandafter \@secondoftwo
 \fi
}%
\providecommand \@ifx [1]{%
 \ifx #1\expandafter \@firstoftwo
 \else \expandafter \@secondoftwo
 \fi
}%
\providecommand \natexlab [1]{#1}%
\providecommand \enquote  [1]{``#1''}%
\providecommand \bibnamefont  [1]{#1}%
\providecommand \bibfnamefont [1]{#1}%
\providecommand \citenamefont [1]{#1}%
\providecommand \href@noop [0]{\@secondoftwo}%
\providecommand \href [0]{\begingroup \@sanitize@url \@href}%
\providecommand \@href[1]{\@@startlink{#1}\@@href}%
\providecommand \@@href[1]{\endgroup#1\@@endlink}%
\providecommand \@sanitize@url [0]{\catcode `\\12\catcode `\$12\catcode
  `\&12\catcode `\#12\catcode `\^12\catcode `\_12\catcode `\%12\relax}%
\providecommand \@@startlink[1]{}%
\providecommand \@@endlink[0]{}%
\providecommand \url  [0]{\begingroup\@sanitize@url \@url }%
\providecommand \@url [1]{\endgroup\@href {#1}{\urlprefix }}%
\providecommand \urlprefix  [0]{URL }%
\providecommand \Eprint [0]{\href }%
\providecommand \doibase [0]{http://dx.doi.org/}%
\providecommand \selectlanguage [0]{\@gobble}%
\providecommand \bibinfo  [0]{\@secondoftwo}%
\providecommand \bibfield  [0]{\@secondoftwo}%
\providecommand \translation [1]{[#1]}%
\providecommand \BibitemOpen [0]{}%
\providecommand \bibitemStop [0]{}%
\providecommand \bibitemNoStop [0]{.\EOS\space}%
\providecommand \EOS [0]{\spacefactor3000\relax}%
\providecommand \BibitemShut  [1]{\csname bibitem#1\endcsname}%
\let\auto@bib@innerbib\@empty
\bibitem [{\citenamefont {Zhou}\ \emph {et~al.}(2013)\citenamefont {Zhou},
  \citenamefont {Gao}, \citenamefont {Zhang}, \citenamefont {Luo},
  \citenamefont {Cao}, \citenamefont {Chen}, \citenamefont {Dai},\ and\
  \citenamefont {Liu}}]{zhou2013vo2}%
  \BibitemOpen
  \bibfield  {author} {\bibinfo {author} {\bibfnamefont {J.}~\bibnamefont
  {Zhou}}, \bibinfo {author} {\bibfnamefont {Y.}~\bibnamefont {Gao}}, \bibinfo
  {author} {\bibfnamefont {Z.}~\bibnamefont {Zhang}}, \bibinfo {author}
  {\bibfnamefont {H.}~\bibnamefont {Luo}}, \bibinfo {author} {\bibfnamefont
  {C.}~\bibnamefont {Cao}}, \bibinfo {author} {\bibfnamefont {Z.}~\bibnamefont
  {Chen}}, \bibinfo {author} {\bibfnamefont {L.}~\bibnamefont {Dai}}, \ and\
  \bibinfo {author} {\bibfnamefont {X.}~\bibnamefont {Liu}},\ }\href@noop {}
  {\bibfield  {journal} {\bibinfo  {journal} {Sci. Rep.}\ }\textbf {\bibinfo
  {volume} {3}},\ \bibinfo {pages} {3029} (\bibinfo {year} {2013})}\BibitemShut
  {NoStop}%
\bibitem [{\citenamefont {Yamamoto}\ \emph {et~al.}(2019)\citenamefont
  {Yamamoto}, \citenamefont {Nouchi}, \citenamefont {Kanki}, \citenamefont
  {Hattori}, \citenamefont {Watanabe}, \citenamefont {Taniguchi}, \citenamefont
  {Ueno},\ and\ \citenamefont {Tanaka}}]{yamamoto2019gate}%
  \BibitemOpen
  \bibfield  {author} {\bibinfo {author} {\bibfnamefont {M.}~\bibnamefont
  {Yamamoto}}, \bibinfo {author} {\bibfnamefont {R.}~\bibnamefont {Nouchi}},
  \bibinfo {author} {\bibfnamefont {T.}~\bibnamefont {Kanki}}, \bibinfo
  {author} {\bibfnamefont {A.~N.}\ \bibnamefont {Hattori}}, \bibinfo {author}
  {\bibfnamefont {K.}~\bibnamefont {Watanabe}}, \bibinfo {author}
  {\bibfnamefont {T.}~\bibnamefont {Taniguchi}}, \bibinfo {author}
  {\bibfnamefont {K.}~\bibnamefont {Ueno}}, \ and\ \bibinfo {author}
  {\bibfnamefont {H.}~\bibnamefont {Tanaka}},\ }\href@noop {} {\bibfield
  {journal} {\bibinfo  {journal} {ACS applied materials \& interfaces}\
  }\textbf {\bibinfo {volume} {11}},\ \bibinfo {pages} {3224} (\bibinfo {year}
  {2019})}\BibitemShut {NoStop}%
\bibitem [{\citenamefont {Brockman}\ \emph {et~al.}(2011)\citenamefont
  {Brockman}, \citenamefont {Aetukuri}, \citenamefont {Topuria}, \citenamefont
  {Samant}, \citenamefont {Roche},\ and\ \citenamefont
  {Parkin}}]{brockman2011increased}%
  \BibitemOpen
  \bibfield  {author} {\bibinfo {author} {\bibfnamefont {J.}~\bibnamefont
  {Brockman}}, \bibinfo {author} {\bibfnamefont {N.}~\bibnamefont {Aetukuri}},
  \bibinfo {author} {\bibfnamefont {T.}~\bibnamefont {Topuria}}, \bibinfo
  {author} {\bibfnamefont {M.}~\bibnamefont {Samant}}, \bibinfo {author}
  {\bibfnamefont {K.}~\bibnamefont {Roche}}, \ and\ \bibinfo {author}
  {\bibfnamefont {S.}~\bibnamefont {Parkin}},\ }\href@noop {} {\bibfield
  {journal} {\bibinfo  {journal} {Applied Physics Letters}\ }\textbf {\bibinfo
  {volume} {98}},\ \bibinfo {pages} {152105} (\bibinfo {year}
  {2011})}\BibitemShut {NoStop}%
\bibitem [{\citenamefont {Mott}\ and\ \citenamefont
  {Friedman}(1974)}]{mott1974metal}%
  \BibitemOpen
  \bibfield  {author} {\bibinfo {author} {\bibfnamefont {N.~F.}\ \bibnamefont
  {Mott}}\ and\ \bibinfo {author} {\bibfnamefont {L.}~\bibnamefont
  {Friedman}},\ }\href@noop {} {\bibfield  {journal} {\bibinfo  {journal}
  {Philosophical Magazine}\ }\textbf {\bibinfo {volume} {30}},\ \bibinfo
  {pages} {389} (\bibinfo {year} {1974})}\BibitemShut {NoStop}%
\bibitem [{\citenamefont {Joshi}\ \emph {et~al.}(2019)\citenamefont {Joshi},
  \citenamefont {Cirino}, \citenamefont {Morley},\ and\ \citenamefont
  {Lederman}}]{PhysRevMaterials.3.124602}%
  \BibitemOpen
  \bibfield  {author} {\bibinfo {author} {\bibfnamefont {T.}~\bibnamefont
  {Joshi}}, \bibinfo {author} {\bibfnamefont {E.}~\bibnamefont {Cirino}},
  \bibinfo {author} {\bibfnamefont {S.~A.}\ \bibnamefont {Morley}}, \ and\
  \bibinfo {author} {\bibfnamefont {D.}~\bibnamefont {Lederman}},\ }\href
  {\doibase 10.1103/PhysRevMaterials.3.124602} {\bibfield  {journal} {\bibinfo
  {journal} {Phys. Rev. Materials}\ }\textbf {\bibinfo {volume} {3}},\ \bibinfo
  {pages} {124602} (\bibinfo {year} {2019})}\BibitemShut {NoStop}%
\bibitem [{\citenamefont {Trastoy}\ \emph
  {et~al.}(2018{\natexlab{a}})\citenamefont {Trastoy}, \citenamefont {Camjayi},
  \citenamefont {del Valle}, \citenamefont {Kalcheim}, \citenamefont
  {Crocombette}, \citenamefont {Villegas}, \citenamefont {Rozenberg},
  \citenamefont {Ravelosona},\ and\ \citenamefont
  {Schuller}}]{trastoy2018magnetic}%
  \BibitemOpen
  \bibfield  {author} {\bibinfo {author} {\bibfnamefont {J.}~\bibnamefont
  {Trastoy}}, \bibinfo {author} {\bibfnamefont {A.}~\bibnamefont {Camjayi}},
  \bibinfo {author} {\bibfnamefont {J.}~\bibnamefont {del Valle}}, \bibinfo
  {author} {\bibfnamefont {Y.}~\bibnamefont {Kalcheim}}, \bibinfo {author}
  {\bibfnamefont {J.-P.}\ \bibnamefont {Crocombette}}, \bibinfo {author}
  {\bibfnamefont {J.}~\bibnamefont {Villegas}}, \bibinfo {author}
  {\bibfnamefont {M.}~\bibnamefont {Rozenberg}}, \bibinfo {author}
  {\bibfnamefont {D.}~\bibnamefont {Ravelosona}}, \ and\ \bibinfo {author}
  {\bibfnamefont {I.~K.}\ \bibnamefont {Schuller}},\ }\href@noop {} {\bibfield
  {journal} {\bibinfo  {journal} {arXiv preprint arXiv:1808.03528}\ } (\bibinfo
  {year} {2018}{\natexlab{a}})}\BibitemShut {NoStop}%
\bibitem [{\citenamefont {Kuwamoto}\ \emph {et~al.}(1980)\citenamefont
  {Kuwamoto}, \citenamefont {Honig},\ and\ \citenamefont
  {Appel}}]{kuwamoto1980electrical}%
  \BibitemOpen
  \bibfield  {author} {\bibinfo {author} {\bibfnamefont {H.}~\bibnamefont
  {Kuwamoto}}, \bibinfo {author} {\bibfnamefont {J.}~\bibnamefont {Honig}}, \
  and\ \bibinfo {author} {\bibfnamefont {J.}~\bibnamefont {Appel}},\
  }\href@noop {} {\bibfield  {journal} {\bibinfo  {journal} {Physical Review
  B}\ }\textbf {\bibinfo {volume} {22}},\ \bibinfo {pages} {2626} (\bibinfo
  {year} {1980})}\BibitemShut {NoStop}%
\bibitem [{\citenamefont {McWhan}\ \emph {et~al.}(1969)\citenamefont {McWhan},
  \citenamefont {Rice},\ and\ \citenamefont {Remeika}}]{mcwhan1969mott}%
  \BibitemOpen
  \bibfield  {author} {\bibinfo {author} {\bibfnamefont {D.}~\bibnamefont
  {McWhan}}, \bibinfo {author} {\bibfnamefont {T.}~\bibnamefont {Rice}}, \ and\
  \bibinfo {author} {\bibfnamefont {J.}~\bibnamefont {Remeika}},\ }\href@noop
  {} {\bibfield  {journal} {\bibinfo  {journal} {Physical Review Letters}\
  }\textbf {\bibinfo {volume} {23}},\ \bibinfo {pages} {1384} (\bibinfo {year}
  {1969})}\BibitemShut {NoStop}%
\bibitem [{\citenamefont {Shin}\ \emph {et~al.}(1995)\citenamefont {Shin},
  \citenamefont {Tezuka}, \citenamefont {Kinoshita}, \citenamefont {Ishii},
  \citenamefont {Kashiwakura}, \citenamefont {Takahashi},\ and\ \citenamefont
  {Suda}}]{shin1995photoemission}%
  \BibitemOpen
  \bibfield  {author} {\bibinfo {author} {\bibfnamefont {S.}~\bibnamefont
  {Shin}}, \bibinfo {author} {\bibfnamefont {Y.}~\bibnamefont {Tezuka}},
  \bibinfo {author} {\bibfnamefont {T.}~\bibnamefont {Kinoshita}}, \bibinfo
  {author} {\bibfnamefont {T.}~\bibnamefont {Ishii}}, \bibinfo {author}
  {\bibfnamefont {T.}~\bibnamefont {Kashiwakura}}, \bibinfo {author}
  {\bibfnamefont {M.}~\bibnamefont {Takahashi}}, \ and\ \bibinfo {author}
  {\bibfnamefont {Y.}~\bibnamefont {Suda}},\ }\href@noop {} {\bibfield
  {journal} {\bibinfo  {journal} {Journal of the Physical Society of Japan}\
  }\textbf {\bibinfo {volume} {64}},\ \bibinfo {pages} {1230} (\bibinfo {year}
  {1995})}\BibitemShut {NoStop}%
\bibitem [{\citenamefont {Lupi}\ \emph {et~al.}(2010)\citenamefont {Lupi},
  \citenamefont {Baldassarre}, \citenamefont {Mansart}, \citenamefont
  {Perucchi}, \citenamefont {Barinov}, \citenamefont {Dudin}, \citenamefont
  {Papalazarou}, \citenamefont {Rodolakis}, \citenamefont {Rueff},
  \citenamefont {Iti{\'{e}}}, \citenamefont {Ravy}, \citenamefont {Nicoletti},
  \citenamefont {Postorino}, \citenamefont {Hansmann}, \citenamefont {Parragh},
  \citenamefont {Toschi}, \citenamefont {Saha-Dasgupta}, \citenamefont
  {Andersen}, \citenamefont {Sangiovanni}, \citenamefont {Held},\ and\
  \citenamefont {Marsi}}]{lupi2010microscopic}%
  \BibitemOpen
  \bibfield  {author} {\bibinfo {author} {\bibfnamefont {S.}~\bibnamefont
  {Lupi}}, \bibinfo {author} {\bibfnamefont {L.}~\bibnamefont {Baldassarre}},
  \bibinfo {author} {\bibfnamefont {B.}~\bibnamefont {Mansart}}, \bibinfo
  {author} {\bibfnamefont {A.}~\bibnamefont {Perucchi}}, \bibinfo {author}
  {\bibfnamefont {A.}~\bibnamefont {Barinov}}, \bibinfo {author} {\bibfnamefont
  {P.}~\bibnamefont {Dudin}}, \bibinfo {author} {\bibfnamefont
  {E.}~\bibnamefont {Papalazarou}}, \bibinfo {author} {\bibfnamefont
  {F.}~\bibnamefont {Rodolakis}}, \bibinfo {author} {\bibfnamefont {J.~P.}\
  \bibnamefont {Rueff}}, \bibinfo {author} {\bibfnamefont {J.~P.}\ \bibnamefont
  {Iti{\'{e}}}}, \bibinfo {author} {\bibfnamefont {S.}~\bibnamefont {Ravy}},
  \bibinfo {author} {\bibfnamefont {D.}~\bibnamefont {Nicoletti}}, \bibinfo
  {author} {\bibfnamefont {P.}~\bibnamefont {Postorino}}, \bibinfo {author}
  {\bibfnamefont {P.}~\bibnamefont {Hansmann}}, \bibinfo {author}
  {\bibfnamefont {N.}~\bibnamefont {Parragh}}, \bibinfo {author} {\bibfnamefont
  {A.}~\bibnamefont {Toschi}}, \bibinfo {author} {\bibfnamefont
  {T.}~\bibnamefont {Saha-Dasgupta}}, \bibinfo {author} {\bibfnamefont {O.~K.}\
  \bibnamefont {Andersen}}, \bibinfo {author} {\bibfnamefont {G.}~\bibnamefont
  {Sangiovanni}}, \bibinfo {author} {\bibfnamefont {K.}~\bibnamefont {Held}}, \
  and\ \bibinfo {author} {\bibfnamefont {M.}~\bibnamefont {Marsi}},\
  }\href@noop {} {\bibfield  {journal} {\bibinfo  {journal} {Nature
  communications}\ }\textbf {\bibinfo {volume} {1}},\ \bibinfo {pages} {1}
  (\bibinfo {year} {2010})}\BibitemShut {NoStop}%
\bibitem [{\citenamefont {Limelette}\ \emph {et~al.}(2003)\citenamefont
  {Limelette}, \citenamefont {Georges}, \citenamefont {J{\'e}rome},
  \citenamefont {Wzietek}, \citenamefont {Metcalf},\ and\ \citenamefont
  {Honig}}]{limelette2003universality}%
  \BibitemOpen
  \bibfield  {author} {\bibinfo {author} {\bibfnamefont {P.}~\bibnamefont
  {Limelette}}, \bibinfo {author} {\bibfnamefont {A.}~\bibnamefont {Georges}},
  \bibinfo {author} {\bibfnamefont {D.}~\bibnamefont {J{\'e}rome}}, \bibinfo
  {author} {\bibfnamefont {P.}~\bibnamefont {Wzietek}}, \bibinfo {author}
  {\bibfnamefont {P.}~\bibnamefont {Metcalf}}, \ and\ \bibinfo {author}
  {\bibfnamefont {J.}~\bibnamefont {Honig}},\ }\href@noop {} {\bibfield
  {journal} {\bibinfo  {journal} {Science}\ }\textbf {\bibinfo {volume}
  {302}},\ \bibinfo {pages} {89} (\bibinfo {year} {2003})}\BibitemShut
  {NoStop}%
\bibitem [{\citenamefont {Wickramaratne}\ \emph {et~al.}(2019)\citenamefont
  {Wickramaratne}, \citenamefont {Bernstein},\ and\ \citenamefont
  {Mazin}}]{wickramaratne2019role}%
  \BibitemOpen
  \bibfield  {author} {\bibinfo {author} {\bibfnamefont {D.}~\bibnamefont
  {Wickramaratne}}, \bibinfo {author} {\bibfnamefont {N.}~\bibnamefont
  {Bernstein}}, \ and\ \bibinfo {author} {\bibfnamefont {I.}~\bibnamefont
  {Mazin}},\ }\href@noop {} {\bibfield  {journal} {\bibinfo  {journal}
  {Physical Review B}\ }\textbf {\bibinfo {volume} {99}},\ \bibinfo {pages}
  {214103} (\bibinfo {year} {2019})}\BibitemShut {NoStop}%
\bibitem [{\citenamefont {Panaccione}\ \emph {et~al.}(2007)\citenamefont
  {Panaccione}, \citenamefont {Sacchi}, \citenamefont {Torelli}, \citenamefont
  {Offi}, \citenamefont {Cautero}, \citenamefont {Sergo}, \citenamefont
  {Fondacaro}, \citenamefont {Henriquet}, \citenamefont {Huotari},
  \citenamefont {Monaco},\ and\ \citenamefont
  {Paolasini}}]{panaccione2007bulk}%
  \BibitemOpen
  \bibfield  {author} {\bibinfo {author} {\bibfnamefont {G.}~\bibnamefont
  {Panaccione}}, \bibinfo {author} {\bibfnamefont {M.}~\bibnamefont {Sacchi}},
  \bibinfo {author} {\bibfnamefont {P.}~\bibnamefont {Torelli}}, \bibinfo
  {author} {\bibfnamefont {F.}~\bibnamefont {Offi}}, \bibinfo {author}
  {\bibfnamefont {G.}~\bibnamefont {Cautero}}, \bibinfo {author} {\bibfnamefont
  {R.}~\bibnamefont {Sergo}}, \bibinfo {author} {\bibfnamefont
  {A.}~\bibnamefont {Fondacaro}}, \bibinfo {author} {\bibfnamefont
  {C.}~\bibnamefont {Henriquet}}, \bibinfo {author} {\bibfnamefont
  {S.}~\bibnamefont {Huotari}}, \bibinfo {author} {\bibfnamefont
  {G.}~\bibnamefont {Monaco}}, \ and\ \bibinfo {author} {\bibfnamefont
  {L.}~\bibnamefont {Paolasini}},\ }\href@noop {} {\bibfield  {journal}
  {\bibinfo  {journal} {Journal of Electron Spectroscopy and Related
  Phenomena}\ }\textbf {\bibinfo {volume} {156}},\ \bibinfo {pages} {64}
  (\bibinfo {year} {2007})}\BibitemShut {NoStop}%
\bibitem [{\citenamefont {Wahila}\ \emph {et~al.}(2020)\citenamefont {Wahila},
  \citenamefont {Quackenbush}, \citenamefont {Sadowski}, \citenamefont
  {Krisponeit}, \citenamefont {Flege}, \citenamefont {Tran}, \citenamefont
  {Ong}, \citenamefont {Schlueter}, \citenamefont {Lee}, \citenamefont {Holtz}
  \emph {et~al.}}]{wahila2020breakdown}%
  \BibitemOpen
  \bibfield  {author} {\bibinfo {author} {\bibfnamefont {M.~J.}\ \bibnamefont
  {Wahila}}, \bibinfo {author} {\bibfnamefont {N.~F.}\ \bibnamefont
  {Quackenbush}}, \bibinfo {author} {\bibfnamefont {J.~T.}\ \bibnamefont
  {Sadowski}}, \bibinfo {author} {\bibfnamefont {J.-O.}\ \bibnamefont
  {Krisponeit}}, \bibinfo {author} {\bibfnamefont {J.~I.}\ \bibnamefont
  {Flege}}, \bibinfo {author} {\bibfnamefont {R.}~\bibnamefont {Tran}},
  \bibinfo {author} {\bibfnamefont {S.~P.}\ \bibnamefont {Ong}}, \bibinfo
  {author} {\bibfnamefont {C.}~\bibnamefont {Schlueter}}, \bibinfo {author}
  {\bibfnamefont {T.-L.}\ \bibnamefont {Lee}}, \bibinfo {author} {\bibfnamefont
  {M.~E.}\ \bibnamefont {Holtz}},  \emph {et~al.},\ }\href@noop {} {\bibfield
  {journal} {\bibinfo  {journal} {arXiv preprint arXiv:2012.05306}\ } (\bibinfo
  {year} {2020})}\BibitemShut {NoStop}%
\bibitem [{\citenamefont {Dupuis}\ \emph {et~al.}(2003)\citenamefont {Dupuis},
  \citenamefont {Haija}, \citenamefont {Richter}, \citenamefont {Kuhlenbeck},\
  and\ \citenamefont {Freund}}]{dupuis2003v2o3}%
  \BibitemOpen
  \bibfield  {author} {\bibinfo {author} {\bibfnamefont {A.-C.}\ \bibnamefont
  {Dupuis}}, \bibinfo {author} {\bibfnamefont {M.~A.}\ \bibnamefont {Haija}},
  \bibinfo {author} {\bibfnamefont {B.}~\bibnamefont {Richter}}, \bibinfo
  {author} {\bibfnamefont {H.}~\bibnamefont {Kuhlenbeck}}, \ and\ \bibinfo
  {author} {\bibfnamefont {H.-J.}\ \bibnamefont {Freund}},\ }\href@noop {}
  {\bibfield  {journal} {\bibinfo  {journal} {Surface science}\ }\textbf
  {\bibinfo {volume} {539}},\ \bibinfo {pages} {99} (\bibinfo {year}
  {2003})}\BibitemShut {NoStop}%
\bibitem [{\citenamefont {Pfuner}\ \emph {et~al.}(2005)\citenamefont {Pfuner},
  \citenamefont {Schoiswohl}, \citenamefont {Sock}, \citenamefont {Surnev},
  \citenamefont {Ramsey},\ and\ \citenamefont {Netzer}}]{pfuner2005metal}%
  \BibitemOpen
  \bibfield  {author} {\bibinfo {author} {\bibfnamefont {F.}~\bibnamefont
  {Pfuner}}, \bibinfo {author} {\bibfnamefont {J.}~\bibnamefont {Schoiswohl}},
  \bibinfo {author} {\bibfnamefont {M.}~\bibnamefont {Sock}}, \bibinfo {author}
  {\bibfnamefont {S.}~\bibnamefont {Surnev}}, \bibinfo {author} {\bibfnamefont
  {M.}~\bibnamefont {Ramsey}}, \ and\ \bibinfo {author} {\bibfnamefont
  {F.}~\bibnamefont {Netzer}},\ }\href@noop {} {\bibfield  {journal} {\bibinfo
  {journal} {Journal of Physics: Condensed Matter}\ }\textbf {\bibinfo {volume}
  {17}},\ \bibinfo {pages} {4035} (\bibinfo {year} {2005})}\BibitemShut
  {NoStop}%
\bibitem [{\citenamefont {Feiten}\ \emph
  {et~al.}(2015{\natexlab{a}})\citenamefont {Feiten}, \citenamefont
  {Kuhlenbeck},\ and\ \citenamefont {Freund}}]{feiten2015surface1}%
  \BibitemOpen
  \bibfield  {author} {\bibinfo {author} {\bibfnamefont {F.~E.}\ \bibnamefont
  {Feiten}}, \bibinfo {author} {\bibfnamefont {H.}~\bibnamefont {Kuhlenbeck}},
  \ and\ \bibinfo {author} {\bibfnamefont {H.-J.}\ \bibnamefont {Freund}},\
  }\href@noop {} {\bibfield  {journal} {\bibinfo  {journal} {The Journal of
  Physical Chemistry C}\ }\textbf {\bibinfo {volume} {119}},\ \bibinfo {pages}
  {22961} (\bibinfo {year} {2015}{\natexlab{a}})}\BibitemShut {NoStop}%
\bibitem [{\citenamefont {Luo}\ \emph {et~al.}(2004)\citenamefont {Luo},
  \citenamefont {Guo},\ and\ \citenamefont {Wang}}]{luo2004thickness}%
  \BibitemOpen
  \bibfield  {author} {\bibinfo {author} {\bibfnamefont {Q.}~\bibnamefont
  {Luo}}, \bibinfo {author} {\bibfnamefont {Q.}~\bibnamefont {Guo}}, \ and\
  \bibinfo {author} {\bibfnamefont {E.}~\bibnamefont {Wang}},\ }\href@noop {}
  {\bibfield  {journal} {\bibinfo  {journal} {Applied physics letters}\
  }\textbf {\bibinfo {volume} {84}},\ \bibinfo {pages} {2337} (\bibinfo {year}
  {2004})}\BibitemShut {NoStop}%
\bibitem [{\citenamefont {Dillemans}\ \emph {et~al.}(2014)\citenamefont
  {Dillemans}, \citenamefont {Smets}, \citenamefont {Lieten}, \citenamefont
  {Menghini}, \citenamefont {Su},\ and\ \citenamefont
  {Locquet}}]{dillemans2014evidence}%
  \BibitemOpen
  \bibfield  {author} {\bibinfo {author} {\bibfnamefont {L.}~\bibnamefont
  {Dillemans}}, \bibinfo {author} {\bibfnamefont {T.}~\bibnamefont {Smets}},
  \bibinfo {author} {\bibfnamefont {R.}~\bibnamefont {Lieten}}, \bibinfo
  {author} {\bibfnamefont {M.}~\bibnamefont {Menghini}}, \bibinfo {author}
  {\bibfnamefont {C.-Y.}\ \bibnamefont {Su}}, \ and\ \bibinfo {author}
  {\bibfnamefont {J.-P.}\ \bibnamefont {Locquet}},\ }\href@noop {} {\bibfield
  {journal} {\bibinfo  {journal} {Applied Physics Letters}\ }\textbf {\bibinfo
  {volume} {104}},\ \bibinfo {pages} {071902} (\bibinfo {year}
  {2014})}\BibitemShut {NoStop}%
\bibitem [{\citenamefont {Grygiel}\ \emph {et~al.}(2007)\citenamefont
  {Grygiel}, \citenamefont {Simon}, \citenamefont {Mercey}, \citenamefont
  {Prellier}, \citenamefont {Fr{\'e}sard},\ and\ \citenamefont
  {Limelette}}]{grygiel2007thickness}%
  \BibitemOpen
  \bibfield  {author} {\bibinfo {author} {\bibfnamefont {C.}~\bibnamefont
  {Grygiel}}, \bibinfo {author} {\bibfnamefont {C.}~\bibnamefont {Simon}},
  \bibinfo {author} {\bibfnamefont {B.}~\bibnamefont {Mercey}}, \bibinfo
  {author} {\bibfnamefont {W.}~\bibnamefont {Prellier}}, \bibinfo {author}
  {\bibfnamefont {R.}~\bibnamefont {Fr{\'e}sard}}, \ and\ \bibinfo {author}
  {\bibfnamefont {P.}~\bibnamefont {Limelette}},\ }\href@noop {} {\bibfield
  {journal} {\bibinfo  {journal} {Applied Physics Letters}\ }\textbf {\bibinfo
  {volume} {91}},\ \bibinfo {pages} {262103} (\bibinfo {year}
  {2007})}\BibitemShut {NoStop}%
\bibitem [{\citenamefont {Schuler}\ \emph {et~al.}(1997)\citenamefont
  {Schuler}, \citenamefont {Klimm}, \citenamefont {Weissmann}, \citenamefont
  {Renner},\ and\ \citenamefont {Horn}}]{schuler1997influence}%
  \BibitemOpen
  \bibfield  {author} {\bibinfo {author} {\bibfnamefont {H.}~\bibnamefont
  {Schuler}}, \bibinfo {author} {\bibfnamefont {S.}~\bibnamefont {Klimm}},
  \bibinfo {author} {\bibfnamefont {G.}~\bibnamefont {Weissmann}}, \bibinfo
  {author} {\bibfnamefont {C.}~\bibnamefont {Renner}}, \ and\ \bibinfo {author}
  {\bibfnamefont {S.}~\bibnamefont {Horn}},\ }\href@noop {} {\bibfield
  {journal} {\bibinfo  {journal} {Thin Solid Films}\ }\textbf {\bibinfo
  {volume} {299}},\ \bibinfo {pages} {119} (\bibinfo {year}
  {1997})}\BibitemShut {NoStop}%
\bibitem [{\citenamefont {Feiten}\ \emph
  {et~al.}(2015{\natexlab{b}})\citenamefont {Feiten}, \citenamefont {Seifert},
  \citenamefont {Paier}, \citenamefont {Kuhlenbeck}, \citenamefont {Winter},
  \citenamefont {Sauer},\ and\ \citenamefont {Freund}}]{feiten2015surface}%
  \BibitemOpen
  \bibfield  {author} {\bibinfo {author} {\bibfnamefont {F.~E.}\ \bibnamefont
  {Feiten}}, \bibinfo {author} {\bibfnamefont {J.}~\bibnamefont {Seifert}},
  \bibinfo {author} {\bibfnamefont {J.}~\bibnamefont {Paier}}, \bibinfo
  {author} {\bibfnamefont {H.}~\bibnamefont {Kuhlenbeck}}, \bibinfo {author}
  {\bibfnamefont {H.}~\bibnamefont {Winter}}, \bibinfo {author} {\bibfnamefont
  {J.}~\bibnamefont {Sauer}}, \ and\ \bibinfo {author} {\bibfnamefont {H.-J.}\
  \bibnamefont {Freund}},\ }\href@noop {} {\bibfield  {journal} {\bibinfo
  {journal} {Physical Review Letters}\ }\textbf {\bibinfo {volume} {114}},\
  \bibinfo {pages} {216101} (\bibinfo {year} {2015}{\natexlab{b}})}\BibitemShut
  {NoStop}%
\bibitem [{\citenamefont {Kresse}\ \emph {et~al.}(2004)\citenamefont {Kresse},
  \citenamefont {Surnev}, \citenamefont {Schoiswohl},\ and\ \citenamefont
  {Netzer}}]{kresse2004v2o3}%
  \BibitemOpen
  \bibfield  {author} {\bibinfo {author} {\bibfnamefont {G.}~\bibnamefont
  {Kresse}}, \bibinfo {author} {\bibfnamefont {S.}~\bibnamefont {Surnev}},
  \bibinfo {author} {\bibfnamefont {J.}~\bibnamefont {Schoiswohl}}, \ and\
  \bibinfo {author} {\bibfnamefont {F.}~\bibnamefont {Netzer}},\ }\href@noop {}
  {\bibfield  {journal} {\bibinfo  {journal} {Surface science}\ }\textbf
  {\bibinfo {volume} {555}},\ \bibinfo {pages} {118} (\bibinfo {year}
  {2004})}\BibitemShut {NoStop}%
\bibitem [{\citenamefont {Meinders}\ \emph {et~al.}(1993)\citenamefont
  {Meinders}, \citenamefont {Eskes},\ and\ \citenamefont
  {Sawatzky}}]{meinders1993spectral}%
  \BibitemOpen
  \bibfield  {author} {\bibinfo {author} {\bibfnamefont {M.}~\bibnamefont
  {Meinders}}, \bibinfo {author} {\bibfnamefont {H.}~\bibnamefont {Eskes}}, \
  and\ \bibinfo {author} {\bibfnamefont {G.}~\bibnamefont {Sawatzky}},\
  }\href@noop {} {\bibfield  {journal} {\bibinfo  {journal} {Physical Review
  B}\ }\textbf {\bibinfo {volume} {48}},\ \bibinfo {pages} {3916} (\bibinfo
  {year} {1993})}\BibitemShut {NoStop}%
\bibitem [{\citenamefont {Kohno}(2019)}]{kohno2019emergence}%
  \BibitemOpen
  \bibfield  {author} {\bibinfo {author} {\bibfnamefont {M.}~\bibnamefont
  {Kohno}},\ }\href@noop {} {\bibfield  {journal} {\bibinfo  {journal}
  {Physical Review B}\ }\textbf {\bibinfo {volume} {100}},\ \bibinfo {pages}
  {235143} (\bibinfo {year} {2019})}\BibitemShut {NoStop}%
\bibitem [{\citenamefont {Rozenberg}\ \emph {et~al.}(1996)\citenamefont
  {Rozenberg}, \citenamefont {Kotliar},\ and\ \citenamefont
  {Kajueter}}]{rozenberg1996transfer}%
  \BibitemOpen
  \bibfield  {author} {\bibinfo {author} {\bibfnamefont {M.}~\bibnamefont
  {Rozenberg}}, \bibinfo {author} {\bibfnamefont {G.}~\bibnamefont {Kotliar}},
  \ and\ \bibinfo {author} {\bibfnamefont {H.}~\bibnamefont {Kajueter}},\
  }\href@noop {} {\bibfield  {journal} {\bibinfo  {journal} {Physical Review
  B}\ }\textbf {\bibinfo {volume} {54}},\ \bibinfo {pages} {8452} (\bibinfo
  {year} {1996})}\BibitemShut {NoStop}%
\bibitem [{\citenamefont {Schoiswohl}\ \emph {et~al.}(2004)\citenamefont
  {Schoiswohl}, \citenamefont {Sock}, \citenamefont {Surnev}, \citenamefont
  {Ramsey}, \citenamefont {Netzer}, \citenamefont {Kresse},\ and\ \citenamefont
  {Andersen}}]{schoiswohl2004v2o3}%
  \BibitemOpen
  \bibfield  {author} {\bibinfo {author} {\bibfnamefont {J.}~\bibnamefont
  {Schoiswohl}}, \bibinfo {author} {\bibfnamefont {M.}~\bibnamefont {Sock}},
  \bibinfo {author} {\bibfnamefont {S.}~\bibnamefont {Surnev}}, \bibinfo
  {author} {\bibfnamefont {M.}~\bibnamefont {Ramsey}}, \bibinfo {author}
  {\bibfnamefont {F.}~\bibnamefont {Netzer}}, \bibinfo {author} {\bibfnamefont
  {G.}~\bibnamefont {Kresse}}, \ and\ \bibinfo {author} {\bibfnamefont {J.~N.}\
  \bibnamefont {Andersen}},\ }\href@noop {} {\bibfield  {journal} {\bibinfo
  {journal} {Surface science}\ }\textbf {\bibinfo {volume} {555}},\ \bibinfo
  {pages} {101} (\bibinfo {year} {2004})}\BibitemShut {NoStop}%
\bibitem [{\citenamefont {Brundle}\ and\ \citenamefont
  {Crist}(2020)}]{brundle2020x}%
  \BibitemOpen
  \bibfield  {author} {\bibinfo {author} {\bibfnamefont {C.~R.}\ \bibnamefont
  {Brundle}}\ and\ \bibinfo {author} {\bibfnamefont {B.~V.}\ \bibnamefont
  {Crist}},\ }\href {\doibase 10.1116/1.5143897} {\bibfield  {journal}
  {\bibinfo  {journal} {Journal of Vacuum Science {\&} Technology A}\ }\textbf
  {\bibinfo {volume} {38}},\ \bibinfo {pages} {041001} (\bibinfo {year}
  {2020})}\BibitemShut {NoStop}%
\bibitem [{\citenamefont {Kundu}\ and\ \citenamefont
  {Menon}(2017)}]{kundu2017thickness}%
  \BibitemOpen
  \bibfield  {author} {\bibinfo {author} {\bibfnamefont {A.~K.}\ \bibnamefont
  {Kundu}}\ and\ \bibinfo {author} {\bibfnamefont {K.~S.}\ \bibnamefont
  {Menon}},\ }\href@noop {} {\bibfield  {journal} {\bibinfo  {journal} {Surface
  Science}\ }\textbf {\bibinfo {volume} {659}},\ \bibinfo {pages} {43}
  (\bibinfo {year} {2017})}\BibitemShut {NoStop}%
\bibitem [{\citenamefont {Mendialdua}\ \emph {et~al.}(1995)\citenamefont
  {Mendialdua}, \citenamefont {Casanova},\ and\ \citenamefont
  {Barbaux}}]{mendialdua1995xps}%
  \BibitemOpen
  \bibfield  {author} {\bibinfo {author} {\bibfnamefont {J.}~\bibnamefont
  {Mendialdua}}, \bibinfo {author} {\bibfnamefont {R.}~\bibnamefont
  {Casanova}}, \ and\ \bibinfo {author} {\bibfnamefont {Y.}~\bibnamefont
  {Barbaux}},\ }\href@noop {} {\bibfield  {journal} {\bibinfo  {journal}
  {Journal of Electron Spectroscopy and Related Phenomena}\ }\textbf {\bibinfo
  {volume} {71}},\ \bibinfo {pages} {249} (\bibinfo {year} {1995})}\BibitemShut
  {NoStop}%
\bibitem [{\citenamefont {Sawatzky}\ and\ \citenamefont
  {Post}(1979)}]{sawatzky1979x}%
  \BibitemOpen
  \bibfield  {author} {\bibinfo {author} {\bibfnamefont {G.}~\bibnamefont
  {Sawatzky}}\ and\ \bibinfo {author} {\bibfnamefont {D.}~\bibnamefont
  {Post}},\ }\href@noop {} {\bibfield  {journal} {\bibinfo  {journal} {Physical
  Review B}\ }\textbf {\bibinfo {volume} {20}},\ \bibinfo {pages} {1546}
  (\bibinfo {year} {1979})}\BibitemShut {NoStop}%
\bibitem [{\citenamefont {Kamakura}\ \emph {et~al.}(2004)\citenamefont
  {Kamakura}, \citenamefont {Taguchi}, \citenamefont {Chainani}, \citenamefont
  {Takata}, \citenamefont {Horiba}, \citenamefont {Yamamoto}, \citenamefont
  {Tamasaku}, \citenamefont {Nishino}, \citenamefont {Miwa}, \citenamefont
  {Ikenaga}, \citenamefont {Awaji}, \citenamefont {Takeuchi}, \citenamefont
  {Ohashi}, \citenamefont {Senba}, \citenamefont {Namatame}, \citenamefont
  {Taniguchi}, \citenamefont {Ishikawa}, \citenamefont {Kobayashi},\ and\
  \citenamefont {Shin}}]{kamakura2004hard}%
  \BibitemOpen
  \bibfield  {author} {\bibinfo {author} {\bibfnamefont {N.}~\bibnamefont
  {Kamakura}}, \bibinfo {author} {\bibfnamefont {M.}~\bibnamefont {Taguchi}},
  \bibinfo {author} {\bibfnamefont {A.}~\bibnamefont {Chainani}}, \bibinfo
  {author} {\bibfnamefont {Y.}~\bibnamefont {Takata}}, \bibinfo {author}
  {\bibfnamefont {K.}~\bibnamefont {Horiba}}, \bibinfo {author} {\bibfnamefont
  {K.}~\bibnamefont {Yamamoto}}, \bibinfo {author} {\bibfnamefont
  {K.}~\bibnamefont {Tamasaku}}, \bibinfo {author} {\bibfnamefont
  {Y.}~\bibnamefont {Nishino}}, \bibinfo {author} {\bibfnamefont
  {D.}~\bibnamefont {Miwa}}, \bibinfo {author} {\bibfnamefont {E.}~\bibnamefont
  {Ikenaga}}, \bibinfo {author} {\bibfnamefont {M.}~\bibnamefont {Awaji}},
  \bibinfo {author} {\bibfnamefont {A.}~\bibnamefont {Takeuchi}}, \bibinfo
  {author} {\bibfnamefont {H.}~\bibnamefont {Ohashi}}, \bibinfo {author}
  {\bibfnamefont {Y.}~\bibnamefont {Senba}}, \bibinfo {author} {\bibfnamefont
  {H.}~\bibnamefont {Namatame}}, \bibinfo {author} {\bibfnamefont
  {M.}~\bibnamefont {Taniguchi}}, \bibinfo {author} {\bibfnamefont
  {T.}~\bibnamefont {Ishikawa}}, \bibinfo {author} {\bibfnamefont
  {K.}~\bibnamefont {Kobayashi}}, \ and\ \bibinfo {author} {\bibfnamefont
  {S.}~\bibnamefont {Shin}},\ }\href@noop {} {\bibfield  {journal} {\bibinfo
  {journal} {EPL (Europhysics Letters)}\ }\textbf {\bibinfo {volume} {68}},\
  \bibinfo {pages} {557} (\bibinfo {year} {2004})}\BibitemShut {NoStop}%
\bibitem [{\citenamefont {Mo}\ \emph {et~al.}(2003)\citenamefont {Mo},
  \citenamefont {Denlinger}, \citenamefont {Kim}, \citenamefont {Park},
  \citenamefont {Allen}, \citenamefont {Sekiyama}, \citenamefont {Yamasaki},
  \citenamefont {Kadono}, \citenamefont {Suga}, \citenamefont {Saitoh},
  \citenamefont {Muro}, \citenamefont {Metcalf}, \citenamefont {Keller},
  \citenamefont {Held}, \citenamefont {Eyert}, \citenamefont {Anisimov},\ and\
  \citenamefont {Vollhardt}}]{mo2003prominent}%
  \BibitemOpen
  \bibfield  {author} {\bibinfo {author} {\bibfnamefont {S.-K.}\ \bibnamefont
  {Mo}}, \bibinfo {author} {\bibfnamefont {J.~D.}\ \bibnamefont {Denlinger}},
  \bibinfo {author} {\bibfnamefont {H.-D.}\ \bibnamefont {Kim}}, \bibinfo
  {author} {\bibfnamefont {J.-H.}\ \bibnamefont {Park}}, \bibinfo {author}
  {\bibfnamefont {J.~W.}\ \bibnamefont {Allen}}, \bibinfo {author}
  {\bibfnamefont {A.}~\bibnamefont {Sekiyama}}, \bibinfo {author}
  {\bibfnamefont {A.}~\bibnamefont {Yamasaki}}, \bibinfo {author}
  {\bibfnamefont {K.}~\bibnamefont {Kadono}}, \bibinfo {author} {\bibfnamefont
  {S.}~\bibnamefont {Suga}}, \bibinfo {author} {\bibfnamefont {Y.}~\bibnamefont
  {Saitoh}}, \bibinfo {author} {\bibfnamefont {T.}~\bibnamefont {Muro}},
  \bibinfo {author} {\bibfnamefont {P.}~\bibnamefont {Metcalf}}, \bibinfo
  {author} {\bibfnamefont {G.}~\bibnamefont {Keller}}, \bibinfo {author}
  {\bibfnamefont {K.}~\bibnamefont {Held}}, \bibinfo {author} {\bibfnamefont
  {V.}~\bibnamefont {Eyert}}, \bibinfo {author} {\bibfnamefont {V.~I.}\
  \bibnamefont {Anisimov}}, \ and\ \bibinfo {author} {\bibfnamefont
  {D.}~\bibnamefont {Vollhardt}},\ }\href@noop {} {\bibfield  {journal}
  {\bibinfo  {journal} {Physical review letters}\ }\textbf {\bibinfo {volume}
  {90}},\ \bibinfo {pages} {186403} (\bibinfo {year} {2003})}\BibitemShut
  {NoStop}%
\bibitem [{\citenamefont {Rodolakis}\ \emph {et~al.}(2009)\citenamefont
  {Rodolakis}, \citenamefont {Mansart}, \citenamefont {Papalazarou},
  \citenamefont {Gorovikov}, \citenamefont {Vilmercati}, \citenamefont
  {Petaccia}, \citenamefont {Goldoni}, \citenamefont {Rueff}, \citenamefont
  {Lupi}, \citenamefont {Metcalf},\ and\ \citenamefont
  {Marsi}}]{rodolakis2009quasiparticles}%
  \BibitemOpen
  \bibfield  {author} {\bibinfo {author} {\bibfnamefont {F.}~\bibnamefont
  {Rodolakis}}, \bibinfo {author} {\bibfnamefont {B.}~\bibnamefont {Mansart}},
  \bibinfo {author} {\bibfnamefont {E.}~\bibnamefont {Papalazarou}}, \bibinfo
  {author} {\bibfnamefont {S.}~\bibnamefont {Gorovikov}}, \bibinfo {author}
  {\bibfnamefont {P.}~\bibnamefont {Vilmercati}}, \bibinfo {author}
  {\bibfnamefont {L.}~\bibnamefont {Petaccia}}, \bibinfo {author}
  {\bibfnamefont {A.}~\bibnamefont {Goldoni}}, \bibinfo {author} {\bibfnamefont
  {J.~P.}\ \bibnamefont {Rueff}}, \bibinfo {author} {\bibfnamefont
  {S.}~\bibnamefont {Lupi}}, \bibinfo {author} {\bibfnamefont {P.}~\bibnamefont
  {Metcalf}}, \ and\ \bibinfo {author} {\bibfnamefont {M.}~\bibnamefont
  {Marsi}},\ }\href@noop {} {\bibfield  {journal} {\bibinfo  {journal}
  {Physical review letters}\ }\textbf {\bibinfo {volume} {102}},\ \bibinfo
  {pages} {066805} (\bibinfo {year} {2009})}\BibitemShut {NoStop}%
\bibitem [{\citenamefont {Borghi}\ \emph {et~al.}(2009)\citenamefont {Borghi},
  \citenamefont {Fabrizio},\ and\ \citenamefont {Tosatti}}]{borghi2009surface}%
  \BibitemOpen
  \bibfield  {author} {\bibinfo {author} {\bibfnamefont {G.}~\bibnamefont
  {Borghi}}, \bibinfo {author} {\bibfnamefont {M.}~\bibnamefont {Fabrizio}}, \
  and\ \bibinfo {author} {\bibfnamefont {E.}~\bibnamefont {Tosatti}},\
  }\href@noop {} {\bibfield  {journal} {\bibinfo  {journal} {Physical review
  letters}\ }\textbf {\bibinfo {volume} {102}},\ \bibinfo {pages} {066806}
  (\bibinfo {year} {2009})}\BibitemShut {NoStop}%
\bibitem [{\citenamefont {Park}\ \emph {et~al.}(2000)\citenamefont {Park},
  \citenamefont {Tjeng}, \citenamefont {Tanaka}, \citenamefont {Allen},
  \citenamefont {Chen}, \citenamefont {Metcalf}, \citenamefont {Honig},
  \citenamefont {de~Groot},\ and\ \citenamefont {Sawatzky}}]{park2000spin}%
  \BibitemOpen
  \bibfield  {author} {\bibinfo {author} {\bibfnamefont {J.-H.}\ \bibnamefont
  {Park}}, \bibinfo {author} {\bibfnamefont {L.}~\bibnamefont {Tjeng}},
  \bibinfo {author} {\bibfnamefont {A.}~\bibnamefont {Tanaka}}, \bibinfo
  {author} {\bibfnamefont {J.}~\bibnamefont {Allen}}, \bibinfo {author}
  {\bibfnamefont {C.}~\bibnamefont {Chen}}, \bibinfo {author} {\bibfnamefont
  {P.}~\bibnamefont {Metcalf}}, \bibinfo {author} {\bibfnamefont
  {J.}~\bibnamefont {Honig}}, \bibinfo {author} {\bibfnamefont
  {F.}~\bibnamefont {de~Groot}}, \ and\ \bibinfo {author} {\bibfnamefont
  {G.}~\bibnamefont {Sawatzky}},\ }\href@noop {} {\bibfield  {journal}
  {\bibinfo  {journal} {Physical Review B}\ }\textbf {\bibinfo {volume} {61}},\
  \bibinfo {pages} {11506} (\bibinfo {year} {2000})}\BibitemShut {NoStop}%
\bibitem [{\citenamefont {Castellani}\ \emph {et~al.}(1978)\citenamefont
  {Castellani}, \citenamefont {Natoli},\ and\ \citenamefont
  {Ranninger}}]{castellani1978magnetic}%
  \BibitemOpen
  \bibfield  {author} {\bibinfo {author} {\bibfnamefont {C.}~\bibnamefont
  {Castellani}}, \bibinfo {author} {\bibfnamefont {C.}~\bibnamefont {Natoli}},
  \ and\ \bibinfo {author} {\bibfnamefont {J.}~\bibnamefont {Ranninger}},\
  }\href@noop {} {\bibfield  {journal} {\bibinfo  {journal} {Physical Review
  B}\ }\textbf {\bibinfo {volume} {18}},\ \bibinfo {pages} {4945} (\bibinfo
  {year} {1978})}\BibitemShut {NoStop}%
\bibitem [{\citenamefont {Deng}\ \emph {et~al.}(2014)\citenamefont {Deng},
  \citenamefont {Sternbach}, \citenamefont {Haule}, \citenamefont {Basov},\
  and\ \citenamefont {Kotliar}}]{deng2014shining}%
  \BibitemOpen
  \bibfield  {author} {\bibinfo {author} {\bibfnamefont {X.}~\bibnamefont
  {Deng}}, \bibinfo {author} {\bibfnamefont {A.}~\bibnamefont {Sternbach}},
  \bibinfo {author} {\bibfnamefont {K.}~\bibnamefont {Haule}}, \bibinfo
  {author} {\bibfnamefont {D.}~\bibnamefont {Basov}}, \ and\ \bibinfo {author}
  {\bibfnamefont {G.}~\bibnamefont {Kotliar}},\ }\href@noop {} {\bibfield
  {journal} {\bibinfo  {journal} {Physical Review Letters}\ }\textbf {\bibinfo
  {volume} {113}},\ \bibinfo {pages} {246404} (\bibinfo {year}
  {2014})}\BibitemShut {NoStop}%
\bibitem [{\citenamefont {Rozenberg}\ \emph {et~al.}(1995)\citenamefont
  {Rozenberg}, \citenamefont {Kotliar}, \citenamefont {Kajueter}, \citenamefont
  {Thomas}, \citenamefont {Rapkine}, \citenamefont {Honig},\ and\ \citenamefont
  {Metcalf}}]{rozenberg1995optical}%
  \BibitemOpen
  \bibfield  {author} {\bibinfo {author} {\bibfnamefont {M.}~\bibnamefont
  {Rozenberg}}, \bibinfo {author} {\bibfnamefont {G.}~\bibnamefont {Kotliar}},
  \bibinfo {author} {\bibfnamefont {H.}~\bibnamefont {Kajueter}}, \bibinfo
  {author} {\bibfnamefont {G.}~\bibnamefont {Thomas}}, \bibinfo {author}
  {\bibfnamefont {D.}~\bibnamefont {Rapkine}}, \bibinfo {author} {\bibfnamefont
  {J.}~\bibnamefont {Honig}}, \ and\ \bibinfo {author} {\bibfnamefont
  {P.}~\bibnamefont {Metcalf}},\ }\href@noop {} {\bibfield  {journal} {\bibinfo
   {journal} {Physical review letters}\ }\textbf {\bibinfo {volume} {75}},\
  \bibinfo {pages} {105} (\bibinfo {year} {1995})}\BibitemShut {NoStop}%
\bibitem [{\citenamefont {Mo}\ \emph {et~al.}(2006)\citenamefont {Mo},
  \citenamefont {Kim}, \citenamefont {Denlinger}, \citenamefont {Allen},
  \citenamefont {Park}, \citenamefont {Sekiyama}, \citenamefont {Yamasaki},
  \citenamefont {Suga}, \citenamefont {Saitoh}, \citenamefont {Muro},\ and\
  \citenamefont {Metcalf}}]{mo2006photoemission}%
  \BibitemOpen
  \bibfield  {author} {\bibinfo {author} {\bibfnamefont {S.-K.}\ \bibnamefont
  {Mo}}, \bibinfo {author} {\bibfnamefont {H.-D.}\ \bibnamefont {Kim}},
  \bibinfo {author} {\bibfnamefont {J.~D.}\ \bibnamefont {Denlinger}}, \bibinfo
  {author} {\bibfnamefont {J.~W.}\ \bibnamefont {Allen}}, \bibinfo {author}
  {\bibfnamefont {J.-H.}\ \bibnamefont {Park}}, \bibinfo {author}
  {\bibfnamefont {A.}~\bibnamefont {Sekiyama}}, \bibinfo {author}
  {\bibfnamefont {A.}~\bibnamefont {Yamasaki}}, \bibinfo {author}
  {\bibfnamefont {S.}~\bibnamefont {Suga}}, \bibinfo {author} {\bibfnamefont
  {Y.}~\bibnamefont {Saitoh}}, \bibinfo {author} {\bibfnamefont
  {T.}~\bibnamefont {Muro}}, \ and\ \bibinfo {author} {\bibfnamefont
  {P.}~\bibnamefont {Metcalf}},\ }\href@noop {} {\bibfield  {journal} {\bibinfo
   {journal} {Physical Review B}\ }\textbf {\bibinfo {volume} {74}},\ \bibinfo
  {pages} {165101} (\bibinfo {year} {2006})}\BibitemShut {NoStop}%
\bibitem [{\citenamefont {Anisimov}\ \emph {et~al.}(2005)\citenamefont
  {Anisimov}, \citenamefont {Kondakov}, \citenamefont {Kozhevnikov},
  \citenamefont {Nekrasov}, \citenamefont {Pchelkina}, \citenamefont {Allen},
  \citenamefont {Mo}, \citenamefont {Kim}, \citenamefont {Metcalf},
  \citenamefont {Suga}, \citenamefont {Sekiyama}, \citenamefont {Keller},
  \citenamefont {Leonov}, \citenamefont {Ren},\ and\ \citenamefont
  {Vollhardt}}]{anisimov2005full}%
  \BibitemOpen
  \bibfield  {author} {\bibinfo {author} {\bibfnamefont {V.~I.}\ \bibnamefont
  {Anisimov}}, \bibinfo {author} {\bibfnamefont {D.~E.}\ \bibnamefont
  {Kondakov}}, \bibinfo {author} {\bibfnamefont {A.~V.}\ \bibnamefont
  {Kozhevnikov}}, \bibinfo {author} {\bibfnamefont {I.~A.}\ \bibnamefont
  {Nekrasov}}, \bibinfo {author} {\bibfnamefont {Z.~V.}\ \bibnamefont
  {Pchelkina}}, \bibinfo {author} {\bibfnamefont {J.~W.}\ \bibnamefont
  {Allen}}, \bibinfo {author} {\bibfnamefont {S.-K.}\ \bibnamefont {Mo}},
  \bibinfo {author} {\bibfnamefont {H.-D.}\ \bibnamefont {Kim}}, \bibinfo
  {author} {\bibfnamefont {P.}~\bibnamefont {Metcalf}}, \bibinfo {author}
  {\bibfnamefont {S.}~\bibnamefont {Suga}}, \bibinfo {author} {\bibfnamefont
  {A.}~\bibnamefont {Sekiyama}}, \bibinfo {author} {\bibfnamefont
  {G.}~\bibnamefont {Keller}}, \bibinfo {author} {\bibfnamefont
  {I.}~\bibnamefont {Leonov}}, \bibinfo {author} {\bibfnamefont
  {X.}~\bibnamefont {Ren}}, \ and\ \bibinfo {author} {\bibfnamefont
  {D.}~\bibnamefont {Vollhardt}},\ }\href@noop {} {\bibfield  {journal}
  {\bibinfo  {journal} {Physical Review B}\ }\textbf {\bibinfo {volume} {71}},\
  \bibinfo {pages} {125119} (\bibinfo {year} {2005})}\BibitemShut {NoStop}%
\bibitem [{\citenamefont {Lechermann}\ \emph {et~al.}(2018)\citenamefont
  {Lechermann}, \citenamefont {Bernstein}, \citenamefont {Mazin},\ and\
  \citenamefont {Valent{\'\i}}}]{lechermann2018uncovering}%
  \BibitemOpen
  \bibfield  {author} {\bibinfo {author} {\bibfnamefont {F.}~\bibnamefont
  {Lechermann}}, \bibinfo {author} {\bibfnamefont {N.}~\bibnamefont
  {Bernstein}}, \bibinfo {author} {\bibfnamefont {I.}~\bibnamefont {Mazin}}, \
  and\ \bibinfo {author} {\bibfnamefont {R.}~\bibnamefont {Valent{\'\i}}},\
  }\href@noop {} {\bibfield  {journal} {\bibinfo  {journal} {Physical review
  letters}\ }\textbf {\bibinfo {volume} {121}},\ \bibinfo {pages} {106401}
  (\bibinfo {year} {2018})}\BibitemShut {NoStop}%
\bibitem [{\citenamefont {Leonov}\ \emph {et~al.}(2015)\citenamefont {Leonov},
  \citenamefont {Anisimov},\ and\ \citenamefont {Vollhardt}}]{leonov2015metal}%
  \BibitemOpen
  \bibfield  {author} {\bibinfo {author} {\bibfnamefont {I.}~\bibnamefont
  {Leonov}}, \bibinfo {author} {\bibfnamefont {V.}~\bibnamefont {Anisimov}}, \
  and\ \bibinfo {author} {\bibfnamefont {D.}~\bibnamefont {Vollhardt}},\
  }\href@noop {} {\bibfield  {journal} {\bibinfo  {journal} {Physical Review
  B}\ }\textbf {\bibinfo {volume} {91}},\ \bibinfo {pages} {195115} (\bibinfo
  {year} {2015})}\BibitemShut {NoStop}%
\bibitem [{\citenamefont {Kundu}\ and\ \citenamefont
  {Menon}(2018)}]{kundu2018structural}%
  \BibitemOpen
  \bibfield  {author} {\bibinfo {author} {\bibfnamefont {A.~K.}\ \bibnamefont
  {Kundu}}\ and\ \bibinfo {author} {\bibfnamefont {K.~S.}\ \bibnamefont
  {Menon}},\ }\href@noop {} {\bibfield  {journal} {\bibinfo  {journal} {Applied
  Surface Science}\ }\textbf {\bibinfo {volume} {456}},\ \bibinfo {pages} {845}
  (\bibinfo {year} {2018})}\BibitemShut {NoStop}%
\bibitem [{\citenamefont {Kim}\ \emph {et~al.}(1998)\citenamefont {Kim},
  \citenamefont {Kumigashira}, \citenamefont {Ashihara}, \citenamefont
  {Takahashi},\ and\ \citenamefont {Ueda}}]{kim1998high}%
  \BibitemOpen
  \bibfield  {author} {\bibinfo {author} {\bibfnamefont {H.-D.}\ \bibnamefont
  {Kim}}, \bibinfo {author} {\bibfnamefont {H.}~\bibnamefont {Kumigashira}},
  \bibinfo {author} {\bibfnamefont {A.}~\bibnamefont {Ashihara}}, \bibinfo
  {author} {\bibfnamefont {T.}~\bibnamefont {Takahashi}}, \ and\ \bibinfo
  {author} {\bibfnamefont {Y.}~\bibnamefont {Ueda}},\ }\href@noop {} {\bibfield
   {journal} {\bibinfo  {journal} {Physical Review B}\ }\textbf {\bibinfo
  {volume} {57}},\ \bibinfo {pages} {1316} (\bibinfo {year}
  {1998})}\BibitemShut {NoStop}%
\bibitem [{\citenamefont {Trastoy}\ \emph
  {et~al.}(2018{\natexlab{b}})\citenamefont {Trastoy}, \citenamefont
  {Kalcheim}, \citenamefont {del Valle}, \citenamefont {Valmianski},\ and\
  \citenamefont {Schuller}}]{trastoy2018enhanced}%
  \BibitemOpen
  \bibfield  {author} {\bibinfo {author} {\bibfnamefont {J.}~\bibnamefont
  {Trastoy}}, \bibinfo {author} {\bibfnamefont {Y.}~\bibnamefont {Kalcheim}},
  \bibinfo {author} {\bibfnamefont {J.}~\bibnamefont {del Valle}}, \bibinfo
  {author} {\bibfnamefont {I.}~\bibnamefont {Valmianski}}, \ and\ \bibinfo
  {author} {\bibfnamefont {I.~K.}\ \bibnamefont {Schuller}},\ }\href@noop {}
  {\bibfield  {journal} {\bibinfo  {journal} {Journal of materials science}\
  }\textbf {\bibinfo {volume} {53}},\ \bibinfo {pages} {9131} (\bibinfo {year}
  {2018}{\natexlab{b}})}\BibitemShut {NoStop}%
\bibitem [{\citenamefont {Lee}\ \emph {et~al.}(2019)\citenamefont {Lee},
  \citenamefont {Wahila}, \citenamefont {Mukherjee}, \citenamefont {Singh},
  \citenamefont {Eustance}, \citenamefont {Regoutz}, \citenamefont {Paik},
  \citenamefont {Boschker}, \citenamefont {Rodolakis}, \citenamefont {Lee},
  \citenamefont {Schlom},\ and\ \citenamefont {Piper}}]{lee2019cooperative}%
  \BibitemOpen
  \bibfield  {author} {\bibinfo {author} {\bibfnamefont {W.-C.}\ \bibnamefont
  {Lee}}, \bibinfo {author} {\bibfnamefont {M.~J.}\ \bibnamefont {Wahila}},
  \bibinfo {author} {\bibfnamefont {S.}~\bibnamefont {Mukherjee}}, \bibinfo
  {author} {\bibfnamefont {C.~N.}\ \bibnamefont {Singh}}, \bibinfo {author}
  {\bibfnamefont {T.}~\bibnamefont {Eustance}}, \bibinfo {author}
  {\bibfnamefont {A.}~\bibnamefont {Regoutz}}, \bibinfo {author} {\bibfnamefont
  {H.}~\bibnamefont {Paik}}, \bibinfo {author} {\bibfnamefont {J.~E.}\
  \bibnamefont {Boschker}}, \bibinfo {author} {\bibfnamefont {F.}~\bibnamefont
  {Rodolakis}}, \bibinfo {author} {\bibfnamefont {T.-L.}\ \bibnamefont {Lee}},
  \bibinfo {author} {\bibfnamefont {D.~G.}\ \bibnamefont {Schlom}}, \ and\
  \bibinfo {author} {\bibfnamefont {L.~F.~J.}\ \bibnamefont {Piper}},\
  }\href@noop {} {\bibfield  {journal} {\bibinfo  {journal} {Journal of Applied
  Physics}\ }\textbf {\bibinfo {volume} {125}},\ \bibinfo {pages} {082539}
  (\bibinfo {year} {2019})}\BibitemShut {NoStop}%
\bibitem [{\citenamefont {Choi}\ \emph {et~al.}(2021)\citenamefont {Choi},
  \citenamefont {Lee}, \citenamefont {Song}, \citenamefont {Kim}, \citenamefont
  {Ju}, \citenamefont {Kim}, \citenamefont {Kim}, \citenamefont {Yoon},
  \citenamefont {Kim}, \citenamefont {Phan}, \citenamefont {Bae},\ and\
  \citenamefont {Park}}]{choi2021correlation}%
  \BibitemOpen
  \bibfield  {author} {\bibinfo {author} {\bibfnamefont {Y.}~\bibnamefont
  {Choi}}, \bibinfo {author} {\bibfnamefont {D.}~\bibnamefont {Lee}}, \bibinfo
  {author} {\bibfnamefont {S.}~\bibnamefont {Song}}, \bibinfo {author}
  {\bibfnamefont {J.}~\bibnamefont {Kim}}, \bibinfo {author} {\bibfnamefont
  {T.-S.}\ \bibnamefont {Ju}}, \bibinfo {author} {\bibfnamefont
  {H.}~\bibnamefont {Kim}}, \bibinfo {author} {\bibfnamefont {J.}~\bibnamefont
  {Kim}}, \bibinfo {author} {\bibfnamefont {S.}~\bibnamefont {Yoon}}, \bibinfo
  {author} {\bibfnamefont {Y.}~\bibnamefont {Kim}}, \bibinfo {author}
  {\bibfnamefont {T.~B.}\ \bibnamefont {Phan}}, \bibinfo {author}
  {\bibfnamefont {J.-S.}\ \bibnamefont {Bae}}, \ and\ \bibinfo {author}
  {\bibfnamefont {S.}~\bibnamefont {Park}},\ }\href@noop {} {\bibfield
  {journal} {\bibinfo  {journal} {Advanced Electronic Materials}\ ,\ \bibinfo
  {pages} {2000874}} (\bibinfo {year} {2021})}\BibitemShut {NoStop}%
\bibitem [{\citenamefont {Mo}\ \emph {et~al.}(2004)\citenamefont {Mo},
  \citenamefont {Kim}, \citenamefont {Allen}, \citenamefont {Gweon},
  \citenamefont {Denlinger}, \citenamefont {Park}, \citenamefont {Sekiyama},
  \citenamefont {Yamasaki}, \citenamefont {Suga}, \citenamefont {Metcalf},\
  and\ \citenamefont {Held}}]{mo2004filling}%
  \BibitemOpen
  \bibfield  {author} {\bibinfo {author} {\bibfnamefont {S.-K.}\ \bibnamefont
  {Mo}}, \bibinfo {author} {\bibfnamefont {H.-D.}\ \bibnamefont {Kim}},
  \bibinfo {author} {\bibfnamefont {J.~W.}\ \bibnamefont {Allen}}, \bibinfo
  {author} {\bibfnamefont {G.-H.}\ \bibnamefont {Gweon}}, \bibinfo {author}
  {\bibfnamefont {J.~D.}\ \bibnamefont {Denlinger}}, \bibinfo {author}
  {\bibfnamefont {J.-H.}\ \bibnamefont {Park}}, \bibinfo {author}
  {\bibfnamefont {A.}~\bibnamefont {Sekiyama}}, \bibinfo {author}
  {\bibfnamefont {A.}~\bibnamefont {Yamasaki}}, \bibinfo {author}
  {\bibfnamefont {S.}~\bibnamefont {Suga}}, \bibinfo {author} {\bibfnamefont
  {P.}~\bibnamefont {Metcalf}}, \ and\ \bibinfo {author} {\bibfnamefont
  {K.}~\bibnamefont {Held}},\ }\href {\doibase 10.1103/PhysRevLett.93.076404}
  {\bibfield  {journal} {\bibinfo  {journal} {Phys. Rev. Lett.}\ }\textbf
  {\bibinfo {volume} {93}},\ \bibinfo {pages} {076404} (\bibinfo {year}
  {2004})}\BibitemShut {NoStop}%
\bibitem [{\citenamefont {Qazilbash}\ \emph {et~al.}(2008)\citenamefont
  {Qazilbash}, \citenamefont {Schafgans}, \citenamefont {Burch}, \citenamefont
  {Yun}, \citenamefont {Chae}, \citenamefont {Kim}, \citenamefont {Kim},\ and\
  \citenamefont {Basov}}]{qazilbash2008electrodynamics}%
  \BibitemOpen
  \bibfield  {author} {\bibinfo {author} {\bibfnamefont {M.~M.}\ \bibnamefont
  {Qazilbash}}, \bibinfo {author} {\bibfnamefont {A.}~\bibnamefont
  {Schafgans}}, \bibinfo {author} {\bibfnamefont {K.}~\bibnamefont {Burch}},
  \bibinfo {author} {\bibfnamefont {S.}~\bibnamefont {Yun}}, \bibinfo {author}
  {\bibfnamefont {B.}~\bibnamefont {Chae}}, \bibinfo {author} {\bibfnamefont
  {B.}~\bibnamefont {Kim}}, \bibinfo {author} {\bibfnamefont {H.-T.}\
  \bibnamefont {Kim}}, \ and\ \bibinfo {author} {\bibfnamefont
  {D.}~\bibnamefont {Basov}},\ }\href@noop {} {\bibfield  {journal} {\bibinfo
  {journal} {Physical Review B}\ }\textbf {\bibinfo {volume} {77}},\ \bibinfo
  {pages} {115121} (\bibinfo {year} {2008})}\BibitemShut {NoStop}%
\bibitem [{\citenamefont {Stewart}\ \emph {et~al.}(2012)\citenamefont
  {Stewart}, \citenamefont {Brownstead}, \citenamefont {Wang}, \citenamefont
  {West}, \citenamefont {Ramirez}, \citenamefont {Qazilbash}, \citenamefont
  {Perkins}, \citenamefont {Schuller},\ and\ \citenamefont
  {Basov}}]{PhysRevB.85.205113}%
  \BibitemOpen
  \bibfield  {author} {\bibinfo {author} {\bibfnamefont {M.~K.}\ \bibnamefont
  {Stewart}}, \bibinfo {author} {\bibfnamefont {D.}~\bibnamefont {Brownstead}},
  \bibinfo {author} {\bibfnamefont {S.}~\bibnamefont {Wang}}, \bibinfo {author}
  {\bibfnamefont {K.~G.}\ \bibnamefont {West}}, \bibinfo {author}
  {\bibfnamefont {J.~G.}\ \bibnamefont {Ramirez}}, \bibinfo {author}
  {\bibfnamefont {M.~M.}\ \bibnamefont {Qazilbash}}, \bibinfo {author}
  {\bibfnamefont {N.~B.}\ \bibnamefont {Perkins}}, \bibinfo {author}
  {\bibfnamefont {I.~K.}\ \bibnamefont {Schuller}}, \ and\ \bibinfo {author}
  {\bibfnamefont {D.~N.}\ \bibnamefont {Basov}},\ }\href {\doibase
  10.1103/PhysRevB.85.205113} {\bibfield  {journal} {\bibinfo  {journal} {Phys.
  Rev. B}\ }\textbf {\bibinfo {volume} {85}},\ \bibinfo {pages} {205113}
  (\bibinfo {year} {2012})}\BibitemShut {NoStop}%
\bibitem [{\citenamefont {Mahatha}\ and\ \citenamefont
  {Menon}(2010)}]{mahatha2010angle}%
  \BibitemOpen
  \bibfield  {author} {\bibinfo {author} {\bibfnamefont {S.}~\bibnamefont
  {Mahatha}}\ and\ \bibinfo {author} {\bibfnamefont {K.~S.}\ \bibnamefont
  {Menon}},\ }\href@noop {} {\bibfield  {journal} {\bibinfo  {journal} {Curr.
  Sci}\ }\textbf {\bibinfo {volume} {98}},\ \bibinfo {pages} {759} (\bibinfo
  {year} {2010})}\BibitemShut {NoStop}%
\end{thebibliography}%


\providecommand{\latin}[1]{#1}
\makeatletter
\providecommand{\doi}
  {\begingroup\let\do\@makeother\dospecials
  \catcode`\{=1 \catcode`\}=2 \doi@aux}
\providecommand{\doi@aux}[1]{\endgroup\texttt{#1}}
\makeatother
\providecommand*\mcitethebibliography{\thebibliography}
\csname @ifundefined\endcsname{endmcitethebibliography}
  {\let\endmcitethebibliography\endthebibliography}{}

\end{document}